\documentclass[12pt]{article}

\ifx\pdfoutput\undefined
\usepackage[dvips,bookmarks]{hyperref}
\else
\usepackage{hyperref}
\fi
\hypersetup{colorlinks=false,bookmarksopen,bookmarksnumbered,citecolor=blue,
   pdfstartview=FitH}

\usepackage{amssymb}
\usepackage{makeidx}
\usepackage{graphicx}
\usepackage{ccaption}
\usepackage{subfigure}
\usepackage{rotating}
\usepackage{lscape}
\usepackage{verbatim}
\usepackage{amsmath, amsthm,amssymb}
\usepackage{mathrsfs}
\usepackage{hypernat}
\usepackage{hyperref}
\usepackage{amsfonts}
\usepackage{dsfont}
\usepackage{bbm}
\usepackage{slashed, tensor}

\usepackage{amsopn}

\parskip=\medskipamount

\usepackage[mathscr]{euscript} 

\newcommand{\de}{{\nabla}}             
\newcommand{\deb}{{\bar{\nabla}}}

\usepackage{amsmath}
\usepackage{amsfonts}
\usepackage{slashed}
\usepackage{youngtab}

\usepackage{breqn}
\usepackage{booktabs}

\usepackage{tikz}
\usetikzlibrary{arrows,shapes}

\usepackage{hyperref}

\newcommand {\cD}{{\cal D}}
\newcommand {\cE}{{\cal E}}

\newcommand {\cL}{{\cal L}}

\newcommand {\cN}{{\cal N}}
\newcommand {\cO}{{\cal O}}

\newcommand {\cX}{{\cal X}}

\newcommand {\cZ}{{\cal Z}}

\def\a{\alpha}
\def\b{\beta}
\def\c{\chi}
\def\d{\delta}
\def\e{\epsilon}
\def\f{\phi}
\def\g{\gamma}

\def\j{\psi}
\def\k{\kappa}
\def\l{\lambda}
\def\m{\mu}

\def\q{\theta}
\def\r{\rho}
\def\s{\sigma}
\def\t{\tau}

\def\x{\xi}
\def\z{\zeta}
\def\D{\Delta}
\def\F{\Phi}
\def\J{\Psi}
\def\L{\Lambda}
\def\O{\Omega}

\def\S{\Sigma}
\def\U{\Upsilon}
\def\X{\Xi}

\def\ri{{\rm i}}
\def\re{{\rm e}}

\newcommand{\gd}{{\dot\g}}
\newcommand{\dd}{{\dot\d}}
\newcommand{\ad}{{\dot{\alpha}}}
\newcommand{\bd}{{\dot{\beta}}}

\newcommand{\sSU}{\mathsf{SU}}
\newcommand{\sSL}{\mathsf{SL}}

\newcommand{\sSO}{\mathsf{SO}}
\newcommand{\sU}{\mathsf{U}}

\newcommand{\dsC}{{\mathbb C}}

\newcommand{\ve}{\varepsilon}
\newcommand{\cDB}{{\bar\cD}}

\newcommand{\pa}{\partial}
\newcommand{\hf}{\frac12}

\newcommand{\vf}{\varphi}

\newcommand{\be}{\begin{equation}}
\newcommand{\ee}{\end{equation}}
\newcommand{\bea}{\begin{eqnarray}}
\newcommand{\eea}{\end{eqnarray}}
\newcommand{\non}{\nonumber}
\newcommand{\ba}{\begin{array}}
\newcommand{\ea}{\end{array}}

\newcommand{\bm}[1]{\mbox{\boldmath$#1$}}

\def\double #1{#1{\hbox{\kern-2pt $#1$}}}

\newcommand{\sba}{{\bar{\s}}}

\newcommand{\bsubeq}{\begin{subequations}}
\newcommand{\esubeq}{\end{subequations}}

\newcommand{\rd}{\mathrm d}
\newcommand{\bi}{{\bar \i}}

\newcommand{\eol}{\notag \\}

\def \bi{\bibitem}

\numberwithin{equation}{section}  
\newcommand{\Db}{{\bar{D}}}

\begin{document}

\begin{titlepage}
\begin{flushright}
February, 2017 \\
\end{flushright}
\vspace{5mm}

\begin{center}
{\Large \bf Goldstino superfields in \mbox{$\bm{\cN=2}$} supergravity }
\\ 
\end{center}

\begin{center}

{\bf Sergei M. Kuzenko${}^{a}$, Ian N. McArthur${}^{a}$ and
\\
Gabriele Tartaglino-Mazzucchelli${}^{b}$
} \\
\vspace{5mm}

\footnotesize{
${}^{a}${\it School of Physics and Astrophysics M013, The University of Western Australia\\
35 Stirling Highway, Crawley W.A. 6009, Australia}}  
~\\
\vspace{2mm}
\footnotesize{
${}^{b}${\it Instituut voor Theoretische Fysica, KU Leuven,\\
Celestijnenlaan 200D, B-3001 Leuven, Belgium}
}
\vspace{2mm}
~\\
\texttt{sergei.kuzenko@uwa.edu.au, ian.mcarthur@uwa.edu.au,
Gabriele.Tartaglino-Mazzucchelli@fys.kuleuven.be}\\
\vspace{2mm}

\end{center}

\begin{abstract}
\baselineskip=14pt
We present off-shell $\cN=2$ supergravity actions, which exhibit 
spontaneously broken local supersymmetry and allow for de Sitter vacua for certain values of the parameters. 
They are obtained by coupling the standard $\cN=2$ supergravity-matter systems
to the Goldstino superfields introduced in arXiv:1105.3001 and arXiv:1607.01277
in the rigid supersymmetric case.
These $\cN=2$ Goldstino superfields include nilpotent chiral and linear 
supermultiplets. We also describe a new reducible $\cN=1$ Goldstino supermultiplet. 
\end{abstract}

\vfill

\vfill
\end{titlepage}

\tableofcontents
\vspace{1cm}
\bigskip\hrule

\renewcommand{\thefootnote}{\arabic{footnote}}
\setcounter{footnote}{0}


\allowdisplaybreaks

\section{Introduction}

Several years ago, two of us introduced a family of constrained  Goldstino superfields \cite{KM}, in terms of which the models for  
spontaneously broken $\cN=2$ Poincar\'e supersymmetry are formulated.
Some of these Goldstino superfields have been generalised to the case of 
$\cN=2$ anti-de Sitter supersymmetry \cite{McArthur}.
The common feature of the constrained $\cN=2$ superfields given in \cite{KM} 
is that their only independent component fields are 
the two Goldstini.
Therefore, if such a Goldstino superfield is coupled to supergravity in order to describe spontaneously broken $\cN=2$ local supersymmetry, it does not bring in any new degrees of freedom, except for making the gravitini  massive
and generating a positive contribution to the cosmological constant, 
in accordance with the super-Higgs effect \cite{VS,VS2,DZ}.
In particular, the absence of scalars is an attractive feature 
for phenomenological applications.

Some of the $\cN=2$ superfield Goldstino
 models given in \cite{KM} have natural $\cN=1$ counterparts \cite{Rocek,IK78,IK82,SW,KTyler}. In particular, the constrained chiral scalar superfield,
 which will be reviewed in section 2.2, 
is the $\cN=2$ analogue of the $\cN=1$ chiral scalar superfield $\f$ 
\cite{Rocek,IK78},
$\bar D_{\dot\alpha} \f =0$, which is
subject to the constraints \cite{Rocek}: 
\begin{subequations} \label{1.1}
\bea 
\f^2&=&0 ~,  \label{1.1a}\\
{f} \f &=& -\frac 14  \f {\bar D}^2 \bar  \f ~, \label{1.1b}
\eea
\end{subequations}
where $f$ is a real parameter of mass dimension $+2$ which characterises 
the supersymmetry breaking scale. 
As was shown by Ro\v{c}ek \cite{Rocek} (see also \cite{KT0} for a recent review),
the Goldstino, which may be identified with $D_\a \f|_{\q=0}$, 
is the only independent component field 
contained in $\f$. In the case of $\cN=1$ supersymmetry, there is an alternative 
superfield approach to describe the Goldstino dynamics, which was advocated in 
\cite{Casalbuoni,KS}. It consists of getting rid of the nonlinear constraint \eqref{1.1b}
and working with a chiral scalar $\cX$, $\bar D_\ad \cX=0$, which is only constrained to be nilpotent, 
\bea
\cX^2=0~.
\eea
Unlike $\f$, the chiral scalar $\cX$ contains an independent auxiliary field in addition 
to the Goldstino. Nevertheless, the former proves to be a function of 
the latter on the mass shell. In practice, the use of $\cX$ 
is somewhat simpler than that of $\f$ from the point of view 
 of its couplings to supergravity and supersymmetric matter. 
 Conceptually, however, the two constrained superfield realisations $\f$ and $\cX$ are completely equivalent \cite{BHKMS} (as long as one deals with low-energy effective actions without higher-derivative terms). In particular, they lead to equivalent couplings
 to supergravity and supersymmetric matter, see  \cite{BHKMS} for the technical details.
The $\cN=2$ superfield analogue of $\cX$  was given in \cite{CDF},
see section \ref{ReducibleChiralScalarGoldstino} below for a review.
We will demonstrate that this realisation is equivalent to the $\cN=2$  chiral scalar 
Goldstino model of \cite{KM}.

In this paper we propose models for spontaneously broken local $\cN=2$ supersymmetry, which are obtained by coupling the standard off-shell supergravity-matter systems to the Goldstino superfields introduced in \cite{KM,CDF}.
Recently, there has been much interest in $\cN=1$ supergravity coupled to nilpotent Goldstino superfields for several reasons. Firstly, such theories
are interesting from the point of view of cosmology due to the possibility 
of describing inflation \cite{ADFS,FKL}. 
Secondly, every Goldstino superfield coupled to supergravity  provides a universal positive contribution to the cosmological constant 
\cite{LR,BFKVP,HY,K15,BHKMS}, unlike the supersymmetric cosmological term 
 \cite{Townsend} which yields
a negative contribution to the cosmological constant. 
The same property is true for the Goldstino brane \cite{BMST}.
Of course, the observation that the coupling of the Volkov-Akulov theory \cite{VA,AV}
to supergravity generates a model-independent positive contribution to the cosmological constant was made long ago in the frameworks of on-shell supergravity \cite{DZ} and  off-shell supergravity \cite{LR}. But it seems that at that time nobody was interested in generating a positive cosmological constant. 
Cosmological model building \cite{ADFS,FKL}
and the so-called de Sitter supergravity \cite{BFKVP} and its extensions 
\cite{KW,SvdWW} have invigorated interest 
in the coupling of nonlinear supersymmetry to supergravity. 
Nonlinear supersymmetries are also intriguing in the context of amplitudes 
\cite{Kallosh:2016qvo}. 
It is also worth recalling that $\cN=2$ supergravity \cite{FvN} 
realised Einstein's dream of unifying electromagnetism and gravity \cite{vN}
by adding a massless complex gravitino to the photon and graviton. 
When $\cN=2$ supergravity is coupled to any of the Goldstino superfields introduced in 
\cite{KM}, the resulting theory describes (in a unitary gauge) 
the Einstein-Maxwell system coupled to a massive complex gravitino. 
Integrating out  the massive gravitino fields leads to a low-energy 
Einstein-Maxwell theory of purely supersymmetric origin. 

This paper is organised as follows. In section 2 we review 
the Goldstino superfields introduced in \cite{KM,CDF} and elaborate on 
their properties and 
the explicit relationships between them.
In section 3 we couple the chiral scalar and analytic Goldstino 
superfields to supergravity and supersymmetric matter. Section 4 is devoted to 
the coupling of the spinor Goldstino superfield to supergravity. 
Several generalisations of our results are given in section 5. 
The main body of the paper is accompanied by four technical appendices. 
Appendix A describes the component content of the nilpotent chiral scalar 
superfield $\F$. Appendix B gives a summary of the $\sSU(2)$ superspace  \cite{Grimm}, while Appendix C briefly introduces $\cN=2$ conformal superspace \cite{Butter4DN=2}.  Finally, Appendix D  discusses nilpotent $\cN=1$
supergravity following and extending \cite{K15}.
Our two-component notation and conventions correspond to \cite{BK}.


\section{Goldstino superfields in Minkowski superspace}

We start by reviewing  some results of \cite{KM} and elaborating on them.

\subsection{Spinor Goldstino superfields}\label{subsection2.1}

The $\cN=2$ analogue of the nonlinear realisation for $\cN=1$ supersymmetry \cite{Zumino:ChiralNLSusy,IK77,IK78,IK82,SW}, 
in which there is a pair of Goldstone fields $\xi^\a_i(x)$ which mix only with themselves under supersymmetry transformation, is based on the coset parametrisation \cite{KM}
\be
g \big(x, \xi_i (x), \bar{\psi}^i (x) \big) = \re^{\ri( - x^a P_a + f^{-1} \xi^{\a}_{i} (x) Q_{\a}^i)} \, \re^{ \ri  f^{-1}\bar{\psi}_{\ad}^i (x) \bar{Q}^{\ad}_i }~ .
\label{2.1}
\ee
This yields the supersymmetry transformations
\begin{subequations}
\bea
\delta \xi^\a_i &=& f \e^\a_i 
- 2 \ri f^{-1} \xi^{\b}_j \bar{\e}^{\bd j} \partial_{\b \bd}  \xi^\a_i~, \\
\delta \bar{\psi}_{\ad}^i &=& f \bar{\e}_{\ad}^i -  2 \ri  f^{-1} \xi^{\b}_j \bar{\e}^{\bd j} \partial_{\b \bd}  \bar{\psi}_{\ad}^i~.
\eea
\end{subequations}
The construction of ${\cal N} = 2$ superfields associated with these Goldstino
fields proceeds as in the ${\cal N}=1$ case, and the resulting superfields $\Xi^\a_i$ and $\bar{\Psi}_{\ad}^i$ satisfy the following set of constraints involving 
the ${\cal N} = 2$  covariant  derivatives $D_A=(\pa_a, D_\a^i ,  \bar D^\ad_i)$:
\begin{subequations} \label{2.3}
\bea
D_{\b}^j \Xi^\a_{i} &=& f \d_{\b}^\a  \delta^j_i 
~,  \label{2.3a}\\
\bar{D}_{\bd j} \Xi^\a_{i} &=&  - 2 \ri f^{-1}\Xi_j^{\b} \partial_{\b \bd} \Xi^\a_{ i}
~,  \label{2.3b} \\
D_{\b}^j \bar{\Psi}_{\ad}^i &=& 0~, \label{antichiral2} \\
\bar{D}_{\bd j} \bar{\Psi}_{\ad}^i &=&f \ve_{\bd \ad} \delta_j^i
- 2 \ri f^{-1}\Xi_j^{\b } \partial_{\b \bd} \bar{\Psi}_{\ad }^i~ .  \label{2.3d}
\eea
\end{subequations}
The constraints
\eqref{2.3a} and \eqref{2.3b}
 were derived for the first time by Wess \cite{Wess84} as a generalisation of the 
 $\cN=1$ construction \cite{SW}. The constraints \eqref{2.3} tell us that 
 $\x^\a_i=\Xi^\a_{i} |_{\q=0}$ and $\bar \j_\ad^i=\bar{\Psi}_{\ad}^i |_{\q=0}$ are the only independent component fields contained in the Goldstino superfields introduced.

The spinor superfields $\X^\a_i$ and  $\bar \J_\ad^i$ provide equivalent 
descriptions of the Goldstini. It may be checked that
the latter is expressed via the former as
\bea
\bar \J_\ad^i =\frac{1}{f^4}D^4 (\bar \X^i_\ad \X^4) 
= \bar \X^i_\ad +O(\X^3)
~, \qquad \Xi^4:=\frac13 \Xi^{ij}\Xi_{ij}=-\frac13 \Xi^{\a\b}\Xi_{\a\b}~, 
\label{2.5}
\eea
which extends the $\cN=1$ result given in \cite{BHKMS}.
Here and below we make use of the following definitions\footnote{We 
point out that the second-order operators $D_{\ij}$, $D_{\a\b}$, 
$\bar D_{ij} $ and $\bar D_{\ad \bd}$ are symmetric in their indices.} 
\begin{subequations}
\bea
D^4 &=& \frac{1}{48} D^{ij} D_{\ij}  =-\frac{1}{48} D^{\a\b}D_{\a\b} ~, \quad 
D_{ij}= D^\a_i D_{\a j} ~, \quad D_{\a\b}=D^i_\a D_{\b i}~,
\\
\bar D^4 &=& \frac{1}{48} {\bar D}^{ij} \bar D_{\ij}  
=-\frac{1}{48} \bar D^{\ad\bd} \bar D_{\ad\bd} ~, \quad 
\bar D_{ij}= \bar D_{\ad i} \bar D^{\ad}_{ j} 
~, \quad \bar D_{\ad\bd}=\bar D_{\ad i} \bar D_\bd^{ i}~.
\eea
\end{subequations}
The composites $\X_{ij} $ and $\X_{\a\b}$  
in eq. \eqref{2.5} are defined similarly.
Note that eq. \eqref{2.5} implies that $\bar{\psi}_\ad^i=\bar{\xi}_\ad^i+\cdots$, where the ellipsis stands for nonlinear terms in $\xi^\a_i$ and $\bar{\xi}_\ad^i$. 

Eq. (\ref{antichiral2}) means that the spinor  superfields 
$\bar{\Psi}_{\ad }^i$ are antichiral and their complex conjugates $\J_{\a i}$ are chiral, 
and so they provide ingredients for an action obtained by integration over the chiral subspace of ${\cal N} = 2$ Minkowski superspace:
\bea 
S_{\rm Goldstino} =- \frac{1}{2f^2} \int \rd^4 x  \rd^4 {\q} \,\J^4
-\frac{1}{2f^2}\int \rd^4 x  \rd^4 \bar{\q} \,\bar \J^4 ~,
\label{2.4}
\eea
where $\J^4:= \frac{1}{3} \J^{ij} \J_{ij}$,  $\J_{ij}:= \J^{\a}_i \J_{\a j}$ and 
$\J^{ij} = \ve^{ik}\ve^{jl} \J_{kl}$.
Making use of \eqref{2.5} allows us to reformulate the Goldstino action \eqref{2.4}
in terms of $\X^\a_i$ and its conjugate:
\bea
S_{\rm Goldstino} =- \frac{1}{f^6} \int \rd^4 x  \rd^4 {\q} \rd^4 \bar \q \, \X^4 \bar \X^4~.
\label{XiAction}
\eea
This action was given for the first time in Ref. \cite{Kandelakis}, 
which  built on the earlier work \cite{Wess84}. At the component level, 
the functionals \eqref{2.4} and \eqref{XiAction} lead to nonlinear actions,
which prove to be equivalent 
to the $\cN=2$ supersymmetric Volkov-Akulov theory \cite{VA,AV}.
To quadratic order in  Goldstini, the action \eqref{XiAction} is 
\bea
S_{\rm Goldstino} = -  \int \rd^4 x \Big( f^2  + \ri \,  \xi^{\a}_{i} 
\overleftrightarrow{\pa_{\a\ad}}\,
\bar{\x}^{\ad i} 
\Big) +O(\x^4) ~.
\label{288}
\eea
The constant term in the integrand \eqref{288} generates a positive (de Sitter) 
contribution to the cosmological constant when the Goldstino superfields  $\X^\a_i$
are coupled to supergravity, see section \ref{section4}. 

In general, $\cN=2$ supersymmetric Goldstino actions contain terms to sixteenth order in the fields. 
The striking feature of the action (\ref{2.4}) is that it is at most 
of eighth order
in the fields $\x^\a_i$ and $\bar \j_\ad^i$, as a consequence of the constraints \eqref{2.3}. 
To quartic order in Goldstini, the component form of the action (\ref{2.4}) is 
\bea
S_{\rm Goldstino} = &-&  \int \rd^4 x \left( \frac12 f^2         + \ri \,  \xi^{\a}_{i} \partial_{\a \ad} \bar{\j}^{\ad i} 
- \frac{1}{4 f^2} \, \xi^{ij} \partial_{\a \ad}\partial^{\ad \a} \bar{\j}_{ij}  -  \frac{1}{4 f^2} \, \xi^{\a \b} \partial_{\a \ad} \partial_{\b \bd} \bar{\j}^{\ad \bd} \right. \nonumber \\
&+& \left. \frac{1}{f^2} \, \xi^{\a i} (\partial_{\a \ad} \xi^{\b j} )
\partial_{\b}{}^{\ad} \bar{\j}_{ij}  
- \frac{1}{f^2}  \, \xi^{\a i} (\partial_{\a \ad} \xi^{\b}_{i} ) \partial_{\b \bd} \bar{\j}^{\ad \bd}  
-   \frac{1}{4f^2}  \xi^{\a \b} \partial_{\b}{}^{\ad} ( (\partial_{\a \ad} \bar{\j}_{\bd}^{i} ) \bar{\j}^{\bd}_{i} ) \right.  \nonumber \\
&+& \left. \frac{1}{2 f^2} \, \xi^{\a \b} ( \partial_{\a \ad} \bar{\j}^{\ad i} ) \partial_{\b \bd} \bar{\j}^{\bd}_i  + \frac{1}{2 f^2} \, \xi^{\a \b} (\partial_{\a \ad} \partial_{\b \bd}
\bar{\j}^{\ad i} ) \bar{\j}^{\bd}_i + \mathrm{c.c.} \right) +\dots
\eea
This action turns into \eqref{288} once $\j^\a_i$ and  $\bar{\psi}_\ad^i$ are expressed in terms of $\x^\a_i$ and  $\bar{\xi}_\ad^i$.


\subsection{Chiral scalar Goldstino superfield}\label{subsection2.2}

The chiral scalar superfield \cite{KM}
\bea
\F:= \J^4 ~, \qquad \bar D^{\ad }_i \F=0
\label{2.7}
\eea
obeys the following nilpotency conditions
\begin{subequations} \label{2.8}
\bea
\F^2&=&0~, \\
\F D_A D_B \F &=&0~,\\
\F D_A D_B D_C \F &=&0~,
\eea
\end{subequations}
as well as the nonlinear relation
\bea
f \F =  \F {\bar D}^4 \bar \F~, 
\label{2.9}
\eea
which is similar to Ro\v{c}ek's constraint \eqref{1.1b}.
It follows from the definition of $\F$, eq. \eqref{2.7}, and from \eqref{2.3d} that 
$D^4 \F$ is nowhere vanishing.

The chiral scalar $\F$ has been defined as the composite superfield \eqref{2.7} constructed from  $\J^\a_i$. It can also be realised as a different composite,
\bea
\F = \frac{1}{f^7}\bar D^4 (\bar \X^4 \X^4) ~,
\label{2.11}
\eea
which is constructed from  $\X^\a_i$ and its conjugate. 
In both realisations,  the relations \eqref{2.8} and \eqref{2.9}
hold identically. On the other hand, if we view $\F$ as a fundamental Goldstino 
superfield, then \eqref{2.8} and \eqref{2.9} must be imposed as constraints. 
In addition, it is necessary to require $D^4 \F$ to be nowhere vanishing. 
These properties guarantee that the two Goldstini, which occur at order $\q^3$,
 are the only independent component fields of $\F$, 
 see Appendix \ref{Appendix0} for the details.
In this approach,
 the spinor Goldstino superfields
can be realised as  composite ones 
constructed from $\F$ and its conjugate. In particular, one finds
\bea
\X^\a_i =- \frac{f}{12} \frac{D^{\a j} D_{ji} \F}{D^4 \F }~.
\label{2.13}
\eea
It is a constructive exercise to check that this composite superfield obeys 
the constraints \eqref{2.3a} and \eqref{2.3b}.

The Goldstino action takes the form
\bea 
S_{\rm Goldstino} =-  \frac{f}{2} \int \rd^4 x  \rd^4 {\q} \,\F
-\frac{f}{2}  \int \rd^4 x  \rd^4 \bar{\q} \,\bar \F
= - \int \rd^4 x \rd^4 \q  \rd^4 \bar{\q} \,  \bar \F \F ~. 
\label{2.12}
\eea


\subsection{Reducible chiral scalar Goldstino superfield}
\label{ReducibleChiralScalarGoldstino}

Instead of working with the Goldstino superfield $\F$, 
which contains only two independent component fields -- the Goldstini --
one can follow a different path, in the spirit 
of the $\cN=1$ constructions advocated in \cite{Casalbuoni,KS}. 
Specifically, 
one can consider a chiral scalar $X$, $\bar D^\ad_i X=0$, which is only required to obey 
the nilpotency constraints
\begin{subequations} \label{2.14}
\bea
X^2&=&0~, \\
X D_A D_B X &=&0~,\\
X D_A D_B D_C X &=&0~, \label{2.14c}
\eea
\end{subequations}
in conjunction with the requirement that $D^4 X$ be nowhere vanishing, 
$D^4 X\neq 0$.
This approach was pursued in \cite{CDF}.\footnote{The chiral scalar $X$ was the only 
novel $\cN=2$ Goldstino superfield introduced in \cite{CDF}, the others had 
been given five years earlier in \cite{KM}.} 
The chiral superfield $X$ contains two independent component fields 
that we identify with the lowest components of the descendants 
$\c^\a_i := -\frac{1}{12} D^{\a j}D_{ij} X$ and $D^4 X$, which
have the obvious, albeit useful,  properties
\begin{subequations}
\bea
X \c^\a_i &=&0~,\\
D_\b^j \c^\a_i &=&\d^\a_\b \d_i^j D^4X~.
\eea
\end{subequations}
The dynamics of this supermultiplet is governed by the action 
\bea 
\widetilde{S}_{\rm Goldstino} = \int \rd^4 x \rd^4 \q  \rd^4 \bar{\q} \,  \bar X X
- f  \int \rd^4 x  \rd^4 {\q} \, X
- f \int \rd^4 x  \rd^4 \bar{\q} \,\bar X
 ~. 
\label{2.15}
\eea
Making use of the constraints \eqref{2.14}, it is not difficult  to derive the following nonlinear representation\footnote{Relation \eqref{repres} has a natural counterpart 
in the case of $\cN=1$  supersymmetry. Given a nilpotent $\cN=1$ chiral superfield $X$, with the properties $\bar D_\ad X =0$ and $X^2 =0$, it can be represented as 
$X = - \c^2 (D^2 X)^{-1}$, where $\c^2 =\c^\a \c_\a$ and 
$\c_\a = D_\a X$.}
for $X$ \cite{CDF}:
\bea
X = \frac{\c^4 } {(D^4X)^3}
~, \qquad \c^4=\frac13\c^{ij}\c_{ij}=-\frac13\c^{\a\b}\c_{\a\b}~,
\label{repres}
\eea
where $ \c^{ij} =\c^{\a i} \c_\a^j $ and $\c_{\a\b}:=\c^i_{\a}\c_{\b i}$.
With this representation for $X$, the constraints \eqref{2.14} hold identically. 

The Goldstino model \eqref{2.15} is equivalent to the one described by the action
\eqref{2.12}. The simplest way to prove this is by extending the $\cN=1$ analysis of \cite{BHKMS} 
to the $\cN=2$ supersymmetric case. The starting point is to notice that if $X$ obeys 
the constraints 
\eqref{2.14}, then $\re^{-\r} X$ also obeys the same constraints, for every chiral
scalar superfield $\r$, $\bar D^\ad_i \r=0$. This freedom may be used to 
represent 
\bea
X =  \re^\r\F ~, 
\label{2.16}
\eea
where $\F$ is the Goldstino superfield described in the previous subsection.
The superfield $\r$ in \eqref{2.16} is defined modulo gauge transformations
of the form 
\bea
\r ~\to ~ \r + \d \r~, \qquad \F \d \r =0~.
\label{2.17}
\eea
We now make use of the representation \eqref{2.16} and vary the action 
\eqref{2.15} with respect to $\r$, which gives
\bea
X \bar D^4 \bar X = f X~.
\eea
We see that on the mass shell the nilpotent chiral superfield $X$, defined by \eqref{2.16}, 
obeys the same constraint as $\F$, eq. \eqref{2.9}. This means that $\r =1$ modulo
the gauge freedom \eqref{2.17}.

It is worth giving an $\cN=2$ extension of one more important result from 
\cite{BHKMS}. The point is that the representation \eqref{2.13} does not require 
$\F$ to obey the nonlinear constraint \eqref{2.9}; only the nilpotency constraints 
\eqref{2.8} are essential. 
In other words, starting from $X$, it turns out that the composite spinor  superfield
\bea
\X^{\a}_i = f \frac{\c^\a_i}{D^4 X}
\label{Xi}
\eea
obeys the constrains \eqref{2.3a} and \eqref{2.3b} (see also \cite{CDF}). 
Now, given $\X^{\a}_i $, we know that eq. \eqref{2.11} defines 
the chiral scalar Goldstino superfield $\F$ subject to the constraints
\eqref{2.8} and \eqref{2.9}. Therefore, we can always represent
\bea
X = \F + \U~, \qquad 
\F = \frac{1}{f^7}\bar D^4 (\bar \X^4 \X^4)~,
\label{2.22}
\eea
for some chiral scalar $\U$ obeying the 
generalised nilpotency condition 
\bea
2\F \U +\U^2 =0~.
\eea
The two component fields of $X$ now belong to the two different chiral superfields
$\F$ and $\U$, of which $\F$ contains the Goldstino and $\U$
 the auxiliary field.

According to the terminology of \cite{BHKMS}, the $\cN=2$ Goldstino superfields described in sections \ref{subsection2.1} and \ref{subsection2.2} are irreducible in the sense that the Goldstini are the only independent component fields of such a superfield,  while the other component fields are simply composites constructed from 
the Goldstini. There also exist reducible Goldstino superfields. They contain 
certain independent auxiliary fields in addition to the Goldstini. Any reducible Goldstino 
superfield may be represented as a sum of an irreducible Goldstino superfield 
and a ``matter'' superfield, which contains the auxiliary fields. The chiral scalar 
$X$ is an example of a reducible Goldstino superfield. It is represented in the form
\eqref{2.22}, where $\F$ is the irreducible Goldstino superfield and $\U$ the matter one.  


\subsection{Analytic Goldstino superfields}\label{subsection2.4}

The superfields $\X^\a_i$, $\bar \J^i_\ad$ and $\F$, which we have 
described in sections \ref{subsection2.1} and \ref{subsection2.2}, 
 are not the only  irreducible Goldstino superfields considered in \cite{KM}. 
 Another Goldstino multiplet introduced in \cite{KM} is a 
complex linear superfield, $H^{ij}$, constructed originally as a composite of the spinor ones. Here we study its properties in more detail using an alternative realisation for 
$H^{ij}$ as a descendant of $\F$.

Our first observation is  that the degrees of freedom of the chiral  scalar $\F$ can be encoded in the following  complex iso-triplet 
\bea
H^{ij}:=\frac{1}{4}D^{ij}\F
\label{2266}
\eea
and its conjugate $\bar H_{ij}=\overline{H^{ij}}=\frac{1}{4}\Db_{ij}\bar\F$.
By construction, $H^{ij}$   satisfies the analyticity  constraints
\bea
D^{(i}_\a H^{jk)}=0
~,~~~
\Db^{(i}_\ad H^{jk)}=0
~,
\label{linear-0}
\eea
which mean that $H^{ij}$ is a $\cN=2$ linear multiplet \cite{BS,SSW}.
Following the modern projective-superspace terminology 
\cite{LR2008,K-lectures}, one may also refer to $H^{ij}$ as a complex $\cO(2)$ multiplet. 

As shown in Appendix \ref{Appendix0}, the chiral scalar $\F$ is expressed in terms of its descendants $\c^\a_i$ and $F$, defined by \eqref{def-chi-F}, 
according to \eqref{A.2d}.
These  are given in terms of $H^{ij}$ as follows:
\bea
\c^\a_i=-\frac{1}{3}D^{\a j}H_{ij}
~,~~~~~~
F=\frac{1}{12}D^{ij}H_{ij}
~.
\eea
As a result, $\F$ turns into a composite superfield constructed from $H^{ij}$.
In particular, the Goldstini $\c^\a_i|_{\q=0}$ can be read off  from $H^{ij}$
by taking its first spinor derivative. Making use of \eqref{A.2b}, we observe that
$H^{ij}$ satisfies the nilpotency constraint
\bea
H^{(ij}H^{kl)}=0
~~~\Longleftrightarrow~~~
H^{ij}H_{kl}
=\frac{2}{3}\d^{i}_{(k}\d^j_{l)}H^2~, ~~~H^2 =\hf H^{ij}H_{ij}~.
\label{2299}
\eea
Moreover, it holds that
\bea
H^{i_1 j_1}H^{i_2 j_2}H^{i_3 j_3}=0~.
\eea
It may be shown that $H^{ij}$ obeys the  following nonlinear constraints 
\begin{subequations}\label{2311}
\bea
\Db^{\ad}_j H^{ij}=
-4\ri\pa^{\a\ad}\frac{H^{ij}D_{\a}^kH_{jk}}{D \cdot H}
~, \qquad D\cdot H = \hf D^{ij} H_{ij}
\label{2.31a}
\eea
and
\bea
fH^{ij}
=
\frac{1}{6}D^{ij}\Big(\frac{\bar D \cdot \bar H}
{D \cdot H} H^2\Big)~,
\label{eq-H-0}
\eea
\end{subequations}
which complete the list of conditions  $H^{ij}$ has to obey in order to be an irreducible Goldstino superfield.
The Goldstino action \eqref{2.12} turns into 
\bea 
S_{\rm Goldstino} =-  \frac{f}{24} \int \rd^4 x  \,\Big(
D^{ij} H_{ij} + \bar D^{ij} \bar H_{ij} \Big)~.
\label{2.32}
\eea
This action is supersymmetric because it is a variant of the $\cN=2$ linear multiplet action proposed by Sohnius \cite{Sohnius}.

It follows from \eqref{2.31a} that the action \eqref{2.32} can be rewritten in the form 
\bea
S_{\rm Goldstino} =-  \frac{f}{12} \int \rd^4 x  \,\Big(
D^{ij} + \bar D^{ij}  \Big) {\mathbb H}_{ij}~, 
\qquad {\mathbb H}^{ij} := \hf (H^{ij} +  \bar H^{ij} )~.
\eea
One may see that the dynamics of the Goldstini can be described using the real linear multiplet ${\mathbb H}^{ij} $, which is an irreducible Goldstino superfield. 
It satisfies a nilpotency condition of degree 3, 
\bea
{\mathbb H}^{(i_1i_2} {\mathbb H}^{i_3 i_4} {\mathbb H}^{i_5 i_6)} =0~.
\eea
If ${\mathbb H}^{ij} $ is used as a fundamental Goldstino superfield, 
the Goldstini may be defined to be proportional to 
$D^{\a j}{\mathbb H}_{ij} |_{\q=0}$. The nonlinear constraint \eqref{eq-H-0}
may be recast in terms of ${\mathbb H}^{ij} $.

We now introduce one more Goldstino superfield
that  is  a real $\cO(4)$ multiplet  associated with 
$H^{ij}$ and $\bar H^{ij}$.
It is defined by 
\bea
L^{ijkl}:=H^{(ij}\bar{H}^{kl)} = 2 {\mathbb H}^{(ij}{\mathbb H}^{kl)} 
\eea
and obeys the analyticity constraints 
\bea
D_\a^{(i} L^{jklm)}=
\Db_\ad^{(i} L^{jklm)}=0
~.
\label{constr-L}
\eea
The second form of the Goldstino action \eqref{2.12} 
may be recast in the alternative form
\bea
S_{\rm Goldstino}=-\frac{1}{5} \int  {\rm d}^4 x \,D^{ ijkl} 
L_{ijkl}
~,~~~~~~
D^{ijkl}:=\frac{1}{16}D^{(ij}\Db^{kl)}
~.
\label{action-L-0}
\eea
This action is $\cN=2$ supersymmetric because it is a variant of the $\cO(4) $
multiplet action  introduced for the first time by
Sohnius, Stelle and West in \cite{SSW}.

To get further insights into the structure of the constrained superfield $L^{ijkl}$, 
which is, by construction, 
defined on $\cN=2$ Minkowski superspace ${\mathbb M}^{4|8}$, it is 
useful to (i) reformulate $L^{ijkl}$ as a holomorphic superfield on 
a superspace with auxiliary bosonic dimensions, 
${\mathbb M}^{4|8} \times {\mathbb C}P^1$, 
which is the most relevant superspace setting for off-shell $\cN=2$ 
supersymmetric theories; 
and (ii)  make use of the modern projective-superspace 
notation \cite{K-lectures}.\footnote{The superspace ${\mathbb M}^{4|8} \times {\mathbb C}P^1$ was originally  introduced by Rosly \cite{Rosly}. It is the superspace setting for
both the harmonic \cite{GIKOS,GIOS} and the projective \cite{KLR,LR-PS}
superspace approaches to $\cN=2$ supersymmetric theories in four dimensions.
 The precise relationship between these approaches is thoroughly discussed in \cite{K98,K-lectures}.}

Let $ v^i \in {\mathbb C}^2 \setminus \{0\}$
be homogeneous coordinates for ${\mathbb C}P^1$. Given a symmetric 
iso-spinor of rank $n$, $T^{i_1 \dots i_n } = T^{(i_1 \dots i_n)}$, we associate with it 
a holomorphic homogeneous polynomial 
$T^{(n)} (v) := v_{i_1} \dots v_{i_n} T^{i_1 \dots i_n}$,  where the superscript ``$n$''
denotes the degree of homogeneity, that is $ T^{(n)} (c\, v )= c^{n}  \, T^{(n)}(v)$,
with $c\in {\mathbb C} \setminus \{  0 \}$. It is clear that $T^{(n)} (v) $
defines a holomorphic tensor field on ${\mathbb C}P^1$. 
If $T^{i_1 \dots i_n } (z) $ is a superfield  constrained by 
\bea
D_\a^{(i_1} T^{i_2 \dots i_{n+1}) } =0 ~, \qquad 
\bar D_\ad^{(i_1} T^{i_2 \dots i_{n+1}) } =0~,
\eea
 it is called an $\cO(n) $ multiplet.\footnote{In case $n$ is even, $n=2m$, 
 one can consistently define real $\cO(2m)$ multiplets which are subject to the reality condition 
 $\overline{T^{i_1 \dots i_{2m} } } = T_{i_1 \dots i_{2m }} 
 = \ve_{i_1 j_1} \dots \ve_{i_{2m} j_{2m} } T^{j_1 \dots j_{2m} }$.
 } 
The holomorphic superfield
$T^{(n)} (z,v) $ on ${\mathbb M}^{4|8} \times {\mathbb C}P^1$, which is 
associated with the $\cO(n)$ multiplet, obeys the analyticity constraints
\bea
D_\a^{(1)}T^{(n)}= 0~, \qquad \Db_\ad^{(1)}T^{(n)}=0~,
\eea
where we have introduced the first-order operators
$D_\a^{(1)}:=v_iD_\a^i$ and $\Db_\ad^{(1)}:=v_i\Db_\ad^i$, 
which anticommute with each other. The $\cO(n)$ multiplets are examples
of the so-called projective multiplets \cite{LR-PS}, see \cite{K-lectures} for a modern review.

The constraints \eqref{linear-0} and \eqref{constr-L} tell us that 
the Goldstino superfields $H^{ij}$ and $L^{ijkl}$ are $\cO(2)$ and $\cO(4)$ multiplets,
respectively.  They can equivalently be described in terms of 
the projective superfields $H^{(2)} (v) =v_i v_j H^{ij}$ 
and $L^{(4)}(v) =v_iv_jv_kv_lL^{ijkl}$.

The Goldstino superfield $L^{(4)}$ 
satisfies the nilpotency constraints
\bsubeq\label{nilpotent-L-1}
\bea
L^{(4)}
L^{(4)}
&=&0
~,
\\
L^{(4)} D_A D_B L^{(4)} &=&0
~,\\
L^{(4)} D_A D_B D_C L^{(4)} &=&0~,
\eea
\esubeq
as well as the nonlinear relation
\bea
f^2L^{(4)} = \frac{1}{4!}
L^{(4)}D^{(4)} \big(\partial^{(-2)}\big)^4 L^{(4)}~,
\label{nilpotent-L-2}
\eea
which follows from eq. \eqref{eq-H-0}.
Here we have introduced the operators
\bea
D^{(4)}:=v_iv_jv_kv_lD^{ijkl}
~,~~~~~~
 \partial^{(-2)} := \frac{1}{(v , u )} u^i \frac{\partial}{\partial v^i} 
~,
\eea
where  $(v , u): = v^i u_i $, and  $u_i$ is an isospinor constrained by the  only requirement $(v,u) \ne 0$ (which means that $v^i $ and $u^i$ are linearly independent).
It is not difficult to see that 
the right-hand side of \eqref{nilpotent-L-2} is independent of $u_i$.
In what follows, given a symmetric iso-spinor 
$T^{i_1 \dots i_n } $, we will  associate with it  not only 
$T^{(n)} = v_{i_1} \dots v_{i_n} T^{i_1 \dots i_n}$, but also the following object
\bea
T^{(-n)}  :=  \frac{1}{(v,u)^n}  u_{i_1} \dots u_{i_n} T^{i_1 \dots i_n}~.
\eea
With this notation, the constraint \eqref{nilpotent-L-2} turns into
\bea
f^2L^{(4)} = 
L^{(4)}D^{(4)}  L^{(-4)}~.
\eea
It is worth pointing out that the constraints \eqref{nilpotent-L-1} and \eqref{nilpotent-L-2}
are quite similar to \eqref{2.8} and \eqref{2.9}.

It may be seen that $L^{(4)}$ is an irreducible Goldstino superfield. 
To demonstrate the equivalence of this description to those discussed earlier, 
we  point out that the following composite real $\cO(4)$ multiplet 
\bea
L^{(4)}:=\frac{1}{f^6}D^{(4)}(\Xi^4\bar\Xi^4)
\label{L-Xi}
\eea
satisfies the nilpotency constraints \eqref{nilpotent-L-1} and \eqref{nilpotent-L-2}.
This relation may be inverted to express $\X^\a_i$ in terms of $L^{(4)}$.

We believe that the Goldstino superfields $H^{ij}$ and $L^{ijkl}$ 
can be generalised to describe spontaneously broken 
 supersymmetry with eight supercharges in  five and six dimensions
where chiral superfields are not defined in the $\sSU(2)$ covariant formalism.


\subsection{Reducible linear Goldstino superfield}

The linear Goldstino superfield \eqref{2266} is constructed from the 
irreducible chiral scalar Goldstino superfield $\F$. Instead of using $\F$, 
we can choose $X$ to define another complex linear superfield, 
\bea
H_X^{ij}:=\frac{1}{4}D^{ij}X~, 
\eea
which is a reducible Goldstino superfield. It satisfies the same  analyticity 
and nilpotency conditions, eqs. \eqref{linear-0} and \eqref{2299}, that $H^{ij}$ does.
However, there is no constraint \eqref{eq-H-0}  in the case of $H_X^{ij}$.

Within the harmonic superspace approach \cite{GIKOS,GIOS}, one deals with $\sSU(2$) 
harmonics $u^+_i$ and $u^-_i $defined by 
\bea
u^-_i :=\overline{u^{+i}} ~, \qquad u^{+i}u^-_i = 1 
\quad \Longleftrightarrow \quad 
{ \Big(u_i{}^- , u_i{}^+ \Big) \in {\sSU(2)}}~.
\eea
They may be related to the isospinors $v^i$ and $u_i$, which we have used in the previous  subsection, as follows:
\bea
v^i ~\to ~ u^{+i} := \frac{v^i}{\sqrt{v^\dagger v}}~, \qquad 
u_i := u^-_i =  \frac{{\bar v}_i}{\sqrt{v^\dagger v}}~,
\eea
with $\bar v_i := \overline{v^i}$. Associated with $H_X^{ij}$ is the analytic superfield
$H_X^{++}= u^+_i u^+_j H_X^{ij}$.

In terms of $H_X^{++}$ and $\bar H_X^{++}$,
the Goldstino action \eqref{2.15} turns into 
\bea
\widetilde{S}_{\rm Goldstino} &=&\int {\rm d}u  \int {\rm d} \z^{(-4)} \, 
 L^{(+4)} ~, \non \\
 L^{(+4)} &=&
\bar H_X^{++} H_X^{++} + f \Big( (\q^+)^2 + (\bar \q^+)^2 \Big)
\Big( H_X^{++} + \bar H_X^{++} \Big)~,
\label{2.49}
\eea
where the integration is over the analytic subspace of harmonic superspace, 
\bea
  {\rm d} \z^{(-4)}: = 
 {\rm d}^4 x 
  \,  (D^{-})^4~, \qquad 
 (D^-)^{4}
 :=
 \frac{1}{16} 
 (\bar{D}^-)^2 (D^-)^2 ~,
 \eea
 and the $u$-integral denotes the integration over the group manifold $\sSU(2)$ defined as in  \cite{GIKOS}. The second term in the analytic Lagrangian \eqref{2.49}
 involves naked Grassmann variables, however the action 
 proves to be supersymmetric
 \cite{DKT}.


\section{Chiral and analytic Goldstino superfields in supergravity}

In this section we couple the chiral scalar ($\F$ and $X$)
and the analytic ($H^{ij}$) Goldstino superfields,
which have been described in the previous section,   
to $\cN=2$ supergravity and supersymmetric matter. 
Since the two chiral realisations 
have been shown to be equivalent,  here we first provide the locally supersymmetric 
extension of $X$ and then explain how to read off the curved analogue of $\F$. 

In this section we make use of  the superspace formulation for $\cN=2$ conformal supergravity, 
which was developed in \cite{KLRT-M1} and employed in \cite{KLRT-M1,K-08}
to construct general off-shell supergravity-matter couplings.\footnote{This formulation 
is often called $\sSU(2)$ superspace, since the corresponding superspace structure group is $\sSL(2,{\mathbb C}) \times \sSU(2)_R$, with $\sSL(2,{\mathbb C}) $
being the universal cover of the Lorentz group $\sSO_0(3,1)$.
There exist two more superspace 
formulations for $\cN=2$ conformal supergravity, 
which are characterised by larger structure groups, specifically: 
(i) the $\sU(2)$ superspace of \cite{Howe} with the structure group  
$\sSL(2,{\mathbb C}) \times \sU(2)_R$, 
where $\sU(2)_R = \sSU(2)_R \times \sU(1)_R$
denotes  the $\cN=2$ $R$-symmetry group; and (ii) the conformal superspace of \cite{Butter4DN=2}, which naturally leads to the superconformal tensor calculus
\cite{deWvHVP,BdeRdeW,deWvHVP2}.
In the latter formulation, the entire $\cN=2$ superconformal algebra is gauged in superspace. The three formulations prove to be equivalent, 
and they are also related to each other in the following sense:
(i) $\sSU(2) $ superspace is a gauged fixed version of $\sU(2)$ superspace
\cite{KLRT-M2}; and (ii) $\sU(2)$ superspace is a gauge fixed version of conformal 
superspace \cite{Butter4DN=2}. The most general off-shell $\cN=2$ supergravity-matter couplings were constructed in $\sSU(2)$ superspace \cite{KLRT-M1,K-08}, 
a few years before the conformal superspace was introduced.
They can uniquely be lifted to $\sU(2) $ superspace \cite{KLRT-M2} and also to conformal superspace \cite{Butter-hyper}. 
For certain applications, $\sSU(2) $ superspace is the simplest formalism to deal with.
We will use the conformal superspace setting in section \ref{section4}.
} 
A brief summary of the corresponding curved superspace geometry is given in Appendix \ref{AppendixA}. 
The reason this superspace geometry 
is suitable to describe $\cN=2$ conformal supergravity is that it is compatible 
with super-Weyl invariance.
The point is that the algebra of covariant derivatives  \eqref{A.3} preserves its 
functional form under the super-Weyl transformations  \cite{KLRT-M1}
\begin{subequations}
\bea
\d_{\s} \cD_\a^i&=&\hf\sba\cD_\a^i+\cD^{\g i}\s \,M_{\g\a}-\cD_{\a k}\s \,J^{ki}~, \\
\d_{\s} \cDB_{\ad i}&=&\hf\s\cDB_{\ad i}+\cDB^{\gd}_{i}\sba \,\bar{M}_{\gd\ad}
+ \cDB_{\ad}^{k}\sba \,J_{ki}~, 
\label{A.7b}
 \\
\d_{\s} \cD_a&=&
\hf(\s+\sba)\cD_a
+{\frac{\ri}4}(\s_a)^\a{}_{\bd} \cD_{\a}^{ k}\s \,\cDB^{\bd}_{ k}
+{\frac{\ri}4}(\s_a)^{\a}{}_\bd \cDB^{\bd}_{ k}\sba \,\cD_{\a}^{ k} \non \\
&&\quad -{\frac12}\cD^b(\s+\sba)M_{ab}
~, 
\eea
\end{subequations}
with  the parameter $\s$ being an arbitrary covariantly chiral superfield, 
${\bar \cD}_{\ad i} \s=0$.
The  dimension-1 components of the torsion transform 
 as follows:
\begin{subequations}
\bea
\d_{\s} S^{ij}&=&\sba S^{ij}-{\frac14}\cD^{\g(i}\cD^{j)}_\g \s~, 
\label{super-Weyl-S} \\
\d_{\s} Y_{\a\b}&=&\sba Y_{\a\b}-{\frac14}\cD^{k}_{(\a}\cD_{\b)k}\s~,
\label{super-Weyl-Y} \\
\d_{\s} {W}_{\a \b}&=&\s {W}_{\a \b }~, \label{A.9c}\\
\d_{\s} G_{\a\bd} &=&
\hf(\s+\sba)G_{\a\bd} -{\frac{\ri}4}
\cD_{\a \bd} (\s-\sba)~.
 \label{super-Weyl-G} 
\eea
\end{subequations}
As is seen from \eqref{A.9c},
the covariantly chiral symmetric spinor $W_{\a\b}$ transforms homogeneously, 
and therefore it is a superfield extension of the Weyl tensor, known 
as  the $\cN=2$ super-Weyl tensor 
\cite{Siegel:1980bp,BdeRdeW,Howe}. 


\subsection{Two realisations for the chiral Goldstino superfield}

The Goldstino superfield $X$ is covariantly chiral, 
\bea
\bar \cD^\ad_i X =0~,
\eea
and obeys the nilpotency constraints 
\begin{subequations} \label{3.4}
\bea
X^2&=&0~, \label{3.4a} \\
X \cD_A \cD_B X &=&0~,\\
X \cD_A \cD_B \cD_C X &=&0~.
\eea
\end{subequations}
We choose $X$ to be inert under the super-Weyl transformations, 
\bea
\d_\s X=0~.
\eea
The  constraints \eqref{3.4} are clearly super-Weyl invariant. 

As in the rigid supersymmetric case, the Goldstino superfield is subject to the additional requirement that  $\D X$ is nowhere vanishing, 
$\D X\neq 0$, so that  $( \D X)^{-1}$ is well defined. Here $\D$ denotes the complex conjugate of the $\cN=2$ chiral projection operator \cite{Muller}
\bea
\bar{\D}
&=&\frac{1}{96} \Big((\cDB^{ij}+16\bar{S}^{ij})\cDB_{ij}
-(\cDB^{\ad\bd}-16\bar{Y}^{\ad\bd})\cDB_{\ad\bd} \Big)
\non\\
&=&\frac{1}{96} \Big(\cDB_{ij}(\cDB^{ij}+16\bar{S}^{ij})
-\cDB_{\ad\bd}(\cDB^{\ad\bd}-16\bar{Y}^{\ad\bd}) \Big)~,
\label{chiral-pr}
\eea
with $\cDB^{\ad\bd}:=\cDB^{(\ad}_k\cDB^{\bd)k}$.
Its main properties can be summarised 
in terms of  an arbitrary super-Weyl inert scalar superfield $U$
as follows:
\begin{subequations} 
\bea
{\bar \cD}^{\ad}_i \bar{\D} U &=&0~, \\
\d_\s U = 0 \quad \Longrightarrow \quad 
\d_\s \bar \D U &=& 2\s \bar \D U~,  \label{2.5b}\\
\int \rd^4 x {\rm d}^4\q {\rm d}^4{\bar \q}\,E\, U
&=& \int {\rm d}^4x {\rm d}^4 \q \, \cE \, \bar{\D} U ~.
\label{chiralproj1} 
\eea
\end{subequations}
Here $E^{-1}= {\rm Ber}(E_A{}^M)$ is the full superspace measure,  
and  $\cE$ denotes the chiral density. 
The derivation of \eqref{chiralproj1} can be found in \cite{KT-M09}.

We postulate the Goldstino superfield action in curved superspace to be 
\bea 
\widetilde{S}_{\rm Goldstino} = \int \rd^4 x \rd^4 \q  \rd^4 \bar{\q}  \,E \,  \bar X X
- \left\{ f  \int \rd^4 x  \rd^4 {\q} \,\cE \, W^2X +{\rm c.c.} \right\} ~,
 \label{3.8}
\eea
as a natural curved-superspace extension of \eqref{2.15}.
Here $W$ denotes the field strength of an Abelian vector multiplet. 
It is a covariantly chiral superfield, $\cDB^\ad_i W= 0$,
which is subject to the 
constraint\footnote{Every covariantly chiral superfield $W$ under 
the additional reality condition \eqref{3.9} is called reduced chiral.}
 \cite{GSW,Howe}
\bea
\S^{ij} := \frac{1}{ 4} \Big(\cD^{ij}+4S^{ij}\Big) W&=&\frac{1}{ 4} 
\Big(\cDB^{ij}+4\bar{S}^{ij}\Big)\bar{W} ~,
\label{3.9}
\eea
and is characterised by the super-Weyl transformation law
\bea
\d_\s W= \s W~.
\eea
It is assumed that $W$ is nowhere vanishing, $W\neq 0$,
and therefore it may be identified with one of the two supergravity 
compensators.\footnote{The choice of the second compensator is not unique.
Different choices lead to different off-shell formulations for $\cN=2$ supergravity
\cite{deWPV}.}
The Goldstino action \eqref{3.8} is super-Weyl invariant. 

The constraints \eqref{3.4} are preserved if $X$ is locally rescaled, 
\bea
X ~ \to ~ \re^\t X ~, \qquad \bar \cD^\ad_i \t =0~,
\label{3.11}
\eea
for an arbitrary covariantly chiral scalar $\t$.  
Requiring the action \eqref{3.8} to be stationary under arbitrary displacements \eqref{3.11} gives 
\bea
X = \F~, \qquad f  W^2 \F = \F \bar \D \bar \F
~.
\label{3.12}
\eea
The constraint on $\F$ is the curved-superspace generalisation of  \eqref{2.9}. 
Making use of \eqref{3.12}, the Goldstino action \eqref{3.8} reduces to
\bea 
S_{\rm Goldstino} =
- \frac{f}{2}  \int \rd^4 x  \rd^4 {\q} \,\cE \, W^2 \F +{\rm c.c.} 
\label{3.13}
\eea


\subsection{Spontaneously broken supergravity}

In this subsection we present two off-shell models for spontaneously 
broken $\cN=2$ supergravity. They are described by actions of the form
\begin{subequations}
\bea
\widetilde{S} = \widetilde{S}_{\rm Goldstino} + S_{\rm SUGRA}~.
\label{3.14}
\eea
Here the Goldstino action is given by \eqref{3.8}, and 
$S_{\rm SUGRA}$ stands for a pure supergravity action. 
Requiring this action to be stationary under arbitrary displacements \eqref{3.11} 
turns $X$ into $\F$ defined by \eqref{3.12}, and the action \eqref{3.14} into 
\bea
{S} = {S}_{\rm Goldstino} + S_{\rm SUGRA}~,
\label{3.14b}
\eea
\end{subequations}
with $S_{\rm Goldstino}$ being given by \eqref{3.13}. 
Below we will consider two different off-shell formulations for $\cN=2$ supergravity.

Let us first consider 
the minimal formulation for $\cN=2$  supergravity with two compensators,
the vector multiplet and the (improved)  tensor multiplet, 
proposed in 1983 by de Wit, Philippe and Van Proeyen
\cite{deWPV}. In superspace, 
the corresponding gauge-invariant supergravity action can be written in the form given 
in \cite{BK11}
\bea
S_{\rm SUGRA}  &=& \frac{1}{ \k^2} \int \rd^4 x {\rm d}^4\q \, \cE \, \Big\{
\J {\mathbb W} - \frac{1}{4} W^2 +m \J W \Big\}          +{\rm c.c.}  
\label{3.15} 
\eea
where $\k$ is the gravitational constant, $m$ the cosmological parameter,
and $\mathbb W $ denotes the following reduced chiral superfield\footnote{This multiplet  was  originally discovered in \cite{deWPV}
using the superconformal tensor calculus.
The regular procedure to derive $\mathbb W $ within the superspace setting
was given in \cite{BK11}.}
\bea
\mathbb W := -\frac{G}{8} (\bar \cD_{ij} + 4 \bar S_{ij}) \left(\frac{G^{ij}}{G^2} \right) ~, 
\qquad G=\sqrt{\frac{1}{2} G^{ij}G_{ij}} ~,
\eea
which is associated with the tensor multiplet. 
The tensor multiplet is usually described using 
its gauge invariant field strength $G^{ij}$,  which is defined to be a  real iso-triplet (that is, 
$G^{ij}=G^{ji}$ and ${\bar G}_{ij}:=\overline{G^{ij} } = G_{ij}$)
subject to the covariant constraints  \cite{BS,SSW}
\bea
\cD^{(i}_\a G^{jk)} =  {\bar \cD}^{(i}_\ad G^{jk)} = 0~,
\label{3177}
\eea
with the super-Weyl transformation law
\bea
\d_\s G^{ij} = (\s +\bar \s) G^{ij}~.
\eea
The constraints \eqref{3177}
are solved  \cite{HST,GS82,Siegel83,Muller86}
in terms of a  covariantly chiral
prepotential $\Psi$, $\bar \cD^\ad_i \J=0$,  as follows:
\begin{align}
\label{eq_Gprepotential}
G^{ij} = \frac{1}{4}\Big( \cD^{ij} +4{S}^{ij}\Big) \Psi
+\frac{1}{4}\Big( \cDB^{ij} +4\bar{S}^{ij}\Big){\bar \Psi}~, \qquad
{\bar \cD}^i_\ad \J=0~.
\end{align}
The field strength $G^{ij}$  is invariant under gauge transformations of the form
\bea
\d_\L \J = {\rm i}\,\L ~, \qquad
\cDB^\ad_i \L= 0~, \quad 
\Big(\cD^{ij}+4S^{ij}\Big) \L=
\Big(\cDB^{ij}+4\bar{S}^{ij}\Big)\bar{\L} ~,~~~
\eea
with $\Lambda$ being an arbitrary  reduced chiral superfield.
The action \eqref{3.15} is invariant under these gauge transformations, 
since both $\mathbb W$ and $W$ are reduced chiral superfields.
The action \eqref{3.15} is also super-Weyl invariant, since the super-Weyl transformation 
laws of $\J$ and $\mathbb W$ are \cite{K-08,BK11}
\bea
\d_\s \J = \s \J~,  \qquad 
\d_\s \mathbb W = \s \mathbb W~.
\eea
Since the iso-vector superfield $G^{ij}$ is one of the two supergravity compensators, 
its length $G$ must be nowhere 
vanishing, $G\neq 0$. 

To vary the action \eqref{3.14} with respect to the vector multiplet, 
it is advantageous to represent $W$ in the form \cite{BK11}
\bea
W = \frac{1}{4}\bar\Delta \Big({\cD}^{ij} + 4 S^{ij}\Big) V_{ij}~,
\eea
where $\bar\Delta$ is the chiral projection operator \eqref{chiral-pr}.
Here the unconstrained real iso-triplet $V_{ij}=V_{ji}$ is 
 the curved-superspace extension 
of Mezincescu's prepotential \cite{Mezincescu} (see also \cite{HST}).
The equation of motion for the vector multiplet is 
\bea
\S^{ij} -m G^{ij} = -2 f\k^2 \big( H_X^{ij} +\bar H_X^{ij} \big)~, 
\qquad H_X^{ij}:= \frac{1}{4} \big({\cD}^{ij} + 4 S^{ij}\big) (WX) ~.
\label{3.20}
\eea
In the limit $f \to 0$,  this equation reduces to the one given in \cite{BK11}. 
Since the tensor multiplet does not couple to the Goldstino superfield in \eqref{3.8}, 
the equation of motion for the tensor multiplet is the same as in pure 
supergravity\footnote{In pure  supergravity, the equation of motion for 
the $\cN=2$ gravitational superfield, which describes the Weyl multiplet,  
is  $G - W \bar W =0$, as demonstrated in  \cite{BK11}.
This equation has a natural counterpart
at the component level  \cite{deWPV}. In the case of spontaneously broken 
supergravity described by the action \eqref{3.14}, this equation gets deformed
by terms involving $X$ and its conjugate, 
$G - W\bar W\big{[}1+2f\k^2 (X+\bar{X}) \big{]} =0$.
}
\cite{BK11}
\bea
\mathbb W + m W &=& 0~.
\eea
Making use of the nilpotency constraint \eqref{3.4a}, from \eqref{3.20} we deduce
\bea
 (\S^{(2)} -m G^{(2)} )^3 =0~,
 \label{3.255}
 \eea
 where $\S^{(2)} (v) = v_i v_j \S^{ij}$ is the real $\cO(2) $ multiplet associated 
with \eqref{3.9}, and $G^{(2)} (v) = v_i v_j G^{ij}$. 
Eq. \eqref{3.255} is a nilpotency condition of degree 3. 
It tells us that 
we are dealing with nilpotent $\cN=2$ supergravity. The equation
\eqref{3.20} is similar to that in spontaneously broken $\cN=1$ supergravity 
\cite{K15,BMST}, see Appendix \ref{appendixD} for a review of 
the construction of \cite{K15}.
%

The supergravity theory \eqref{3.15} possesses a dual formulation 
in which the tensor multiplet compensator is dualised into a polar hypermultiplet 
compensator \cite{K-08}.
To obtain the dual formulation, the first step is to recast the chiral action \eqref{3.15} 
as a projective action. Within the off-shell formulation for general supergravity-matter
systems developed in \cite{KLRT-M1}, a universal locally supersymmetric action 
is given by 
\bea
S &=&
\frac{1}{2\pi} \oint (v, \rd v)
\int \rd^4 x {\rm d}^4\q {\rm d}^4{\bar \q}
\,E\, \frac{{ W}{\bar { W}}\cL^{(2)}}{({ \S}^{(2)})^2}~. 
\label{InvarAc}
\eea
The Lagrangian $\cL^{(2)}(v)$ in \eqref{InvarAc}
is a covariant projective multiplet of weight two, which is real with respect to the 
so-called smile conjugation, 
see \cite{KLRT-M1} for the details. 
The projective Lagrangian corresponding to \eqref{InvarAc} was given in  \cite{K-08}.
It is 
\bea
\k^2 \cL^{(2)}_{\rm SUGRA} =      
 {G}^{(2)}  \ln \frac{{ G}^{(2)}}{{\rm i} \U^{(1)} \breve{\U}{}^{(1)}}      
- \hf {V}{\S}^{(2)} 
+mV G^{(2)}~,
\label{3.27ana}
\eea
where $V (v) $ is the tropical prepotential for the vector multiplet, 
and $\U^{(1)} (v)$ is a weight-one arctic multiplet
(both $\U^{(1)}$ and its smile-conjugate $\breve{\U}{}^{(1)}$ are pure gauge degrees of freedom). The chiral field strength $W$ is constructed in terms 
of the tropical prepotential as follows \cite{KT-M08}:
\bea
W  &=& \frac{1}{8\pi}  \oint (v, {\rm d}v )
 \Big( {\bar \cD}^{(-2)} +4 \bar{S}^{(-2)}\Big) V(v)~, 
\eea
where we have used the notation defined in subsection \ref{subsection2.4}.
This field strength is invariant under gauge transformations 
 of the form:
\bea
\d_\l V = \l + \breve \l ~,
\label{3.29gauge}
\eea
with the gauge parameter $\l(v)$ being a covariant weight-zero arctic multiplet, 
and $\breve{\l}$ its smile-conjugate. 

Unlike the action \eqref{3.15}, the tensor multiplet appears in \eqref{3.27ana}
only via its gauge invariant field strength $G^{(2)}$. 
With reference to the vector multiplet, 
it appears in  \eqref{3.15} only via its gauge invariant field strength, while 
the projective Lagrangian \eqref{3.27ana} involves the gauge prepotential $V$.
The locally supersymmetric action generated by \eqref{3.27ana} 
is invariant under the gauge transformations \eqref{3.29gauge}.

Since the tensor multiplet compensator appears in the Lagrangian \eqref{3.27ana}
only via the gauge invariant field strength, $G^{(2)}$, the tensor multiplet 
can be dualised into an off-shell  polar hypermultiplet following the scheme described in  
\cite{K-08}.  The dual supergravity Lagrangian is
\bea
\k^2 \cL^{(2)}_{\rm SUGRA, dual} = - \hf {V}{\S}^{(2)} 
 -{\rm i} \breve{ \U}^{(1)} {\rm e}^{m V}{ \U}^{(1)}~.
\eea
Under the gauge transformation \eqref{3.29gauge}, the hypermultiplet varies 
as $\d_\l \U^{(1)} = - m \l \U^{(1)}$ such that the supergravity action is gauge invariant.


\subsection{Matter couplings}

General matter couplings for the Goldstini are obtained by replacing the Goldstino action 
\eqref{3.8} with 
\bea 
\widetilde{S}_{\rm Goldstino} ~~\longrightarrow ~~
 \int \rd^4 x \rd^4 \q  \rd^4 \bar{\q}  \,E \,  \cN \bar X X
- \left\{  \int \rd^4 x  \rd^4 {\q} \,\cE \, \cZ X +{\rm c.c.} \right\}~.
\label{3.28}
\eea
Here $\cN$ is a super-Weyl invariant real scalar, $\d_\s \cN=0$, 
while $\cZ$ is a covariantly chiral scalar, $\bar \cD^\ad_i \cZ =0$, 
with the super-Weyl transformation law $\d_\s \cZ = 2\s \cZ$. If the supersymmetric 
matter consists of a set of Abelian vector multiplets described by 
covariantly  chiral field strengths $W_I$,  
then $\cN = \cN (W_I, \bar W_J )$ and $\cZ = \cZ(W_I)$. 
In order to guarantee the required super-Weyl transformation laws, 
the composites $ \cN (W_I, \bar W_J )$ and $ \cZ(W_I)$ must be assigned
certain homogeneity properties. Some of the reduced chiral superfields $W_I$ 
may be composite. For instance, we may choose
\bea
\cZ = W {\frak W} ~, \qquad {\frak W} =  \frac{1}{8\pi}  \oint (v, {\rm d}v )
 \Big( {\bar \cD}^{(-2)} +4 \bar{S}^{(-2)}\Big) K(\U^a, \breve{\U}^{\bar b})~,
\label{3322}
\eea
where $W$ is the chiral compensator, and $K(\vf^a, {\bar \vf}^{\bar b}) $ 
is the K\"ahler potential of a real analytic K\"ahler manifold,
with $a, \bar b=1,\dots, n$. The hypermultiplet variables  $\U^a(v) $ in \eqref{3322}
are covariant  weight-zero arctic multiplets, and 
$\breve{\U}^{\bar b}$ denotes the smile conjugate of $\U^b$, 
see   \cite{KLRT-M1} for the technical details.
The reduced chiral 
superfield $\frak W$ is invariant under K\"ahler transformations  \cite{KT-M08}
\bea 
K(\U, \breve{\U}) ~\to ~ K(\U, \breve{\U}) + \L (\U) +\bar \L (\breve{\U} )~,
\eea
with $\L (\vf) $ being a holomorphic function. The action \eqref{3.28} with $\cZ$ given 
by \eqref{3322} describes the off-shell coupling of the Goldstino superfield to 
hypermultiplets.

Requiring the action \eqref{3.28} to be stationary under arbitrary displacements \eqref{3.11} gives 
\bea
X = \F~, \qquad   \cZ \F = \F \bar \D(\cN  \bar \F)~,
\eea
which is a consistent deformation of the constraint \eqref{3.12}.


\subsection{Analytic Goldstino superfields}

We now recast the Goldstino action \eqref{3.13} in terms of a linear Goldstino superfield.  
This is achieved by recalling the observation   \cite{K-08,KT-M09} that 
the $\cN=2$ locally supersymmetric chiral action 
\bea
S_{\rm chiral}= \int \rd^4 x {\rm d}^4\q \, \cE \, \cL_{\rm c} &+& {\rm c.c.}~,
\qquad {\bar \cD}^\ad_i \cL_{\rm c} =0 ~, 
\eea
can be realised as a projective superspace action:
\bea
S_{\rm chiral}&=&\frac{1}{2\pi} \oint (v,\rd v)
\int \rd^4 x {\rm d}^4\q {\rm d}^4{\bar \q}\, E\,
\frac{{ W}{\bar { W}}  \cL^{(2)}_{\rm c} }{({ \S}^{(2)})^2 }~, \non \\
\cL^{(2)}_{\rm c} &=&
 \frac{1}{4} { V} \,\Big\{ \big( \cD^{(2)} +4{S}^{(2)}\big) \frac{\cL_{\rm c}}{ W}
+\big( \cDB^{(2)} + 4\bar{S}^{(2)}\big)
\frac{{\bar \cL}_{\rm c} }{\bar { W}} \Big\}~,~~~~~~
\eea
where $V$ is the  tropical prepotential for the vector multiplet with field strength $W$. 
Applying this general result to the Goldstino action \eqref{3.13} gives
the projective Lagrangian
\bea
\cL^{(2)}_{\rm Goldstino} =-\frac{f}{2}  V\big( H^{(2)} + \bar H^{(2)} \big) 
\equiv -f V {\mathbb H}^{(2)}~,
\eea
where we have introduced the complex linear Goldstino superfield
\bea
H^{(2)} = v_i v_j H^{ij}~, \qquad H^{ij} = \frac{1}{4} (\cD^{ij} +4S^{ij}) (W\F)~.
\eea
It obeys the analyticity constraints 
\bea
\cD^{(1)}_\a H^{(2)} =  {\bar \cD}^{(1)}_\ad H^{(2)} = 0 \quad \Longleftrightarrow \quad 
\cD^{(i}_\a H^{jk)} =  {\bar \cD}^{(i}_\ad H^{jk)} = 0~,
\eea
as well as the nilpotency condition
\bea
H^{(ij} H^{kl)} =0 \quad \Longleftrightarrow \quad (H^{(2)})^2=0~.
\eea
In terms of the real linear superfield ${\mathbb H}^{(2)}$, the nilpotency condition is 
\bea
({\mathbb H}^{(2)})^3=0~.
\eea

We now show how the spontaneously broken supergravity  \eqref{3.14}
can be reformulated as a nilpotent $\cN=2$ supergravity  theory. 
Varying the action \eqref{3.14} with respect to the tensor multiplet compensator
leads to the equation \eqref{3.20}. We then require the action \eqref{3.14} to be stationary under arbitrary displacements \eqref{3.11}, which implies that  $X=\F$ 
and the action \eqref{3.14} turns into \eqref{3.14b}. 
Then, the equation \eqref{3.20} takes the form 
\bea
\S^{(2)} -m G^{(2)} = -4 f\k^2 {\mathbb H}^{(2)}  ~.
\eea
As a result, the projective Lagrangian for the theory \eqref{3.14b} 
can be written as 
\bea
\k^2 \cL^{(2)}=      
 {G}^{(2)}  \ln \frac{{ G}^{(2)}}{{\rm i} \U^{(1)} \breve{\U}{}^{(1)}}      
- \frac14 {V}{\S}^{(2)} 
+ \frac 34 m V G^{(2)}~.
\eea
This has the form of the supergravity Lagrangian \eqref{3.27ana} with rescaled parameters. 
The two conformal compensators have to obey the nilpotency condition \eqref{3.255}
as well as curved-superspace analogues of the nonlinear constraints \eqref{2311}.


\section{Spinor Goldstino superfields in supergravity}\label{section4}

In this section, we provide a curved-superspace extension of the spinor 
Goldstino superfield $\X^\a_i$ defined by the constraints \eqref{2.3a} and \eqref{2.3b}. 
It is known that some superfield representations of Poincar\'e supersymmetry 
cannot be lifted to the locally supersymmetric case. In particular, 
covariantly chiral $\cN=2$ superfields with Lorentz and $\sSU(2)$
indices cannot be defined in general. 
This means that there is no straightforward curved-superspace
generalisation of the spinor Goldstino superfield $\bar \J_\ad^i$
defined by the constraints \eqref{antichiral2} and \eqref{2.3d}.
Fortunately, the constraints on $\X^\a_i$ allow for a supergravity analogue. 

In order to lift $\X^\a_i$ to supergravity, 
it is advantageous to 
employ the superspace formulation
for $\cN=2$ conformal supergravity developed by Butter\footnote{Conformal 
superspace was originally constructed for 4D $\cN=1$ supergravity \cite{Butter4DN=1}
and then extended to the 4D $\cN=2$ \cite{Butter4DN=2} case,
3D $\cN$-extended conformal supergravity  \cite{BKNT-M1}, 
5D conformal supergravity \cite{BKNT-M5D}
and recently to the  6D $\cN=(1,0)$ case \cite{BKNT}. 
}
\cite{Butter4DN=2}
and further elaborated in \cite{BN}.
We denote by $\de_A=(\de_a,\de^i_\a,\deb^\ad_i)$
the corresponding superspace covariant derivatives. 
Throughout this section we use
the notation and various results
 from Ref. \cite{BN}.
 Appendix \ref{AppendixConfSuperspace} includes 
 those technical details on conformal superspace which are relevant for our analysis.

We proceed by lifting the Goldstino superfield $X$, 
which  has so far been defined in $\sSU(2)$ superspace, 
to conformal superspace. Such a reformulation is unique if $X$
is required to be primary, in addition to being covariantly chiral.
 Then the superconformal properties of $X$ are:
 \bea
{ \mathbb D} X=0~,~~~~~~
YX=0~,~~~~~~
K^AX=0~,~~~~~~
\deb^\ad_i X=0
~,
 \eea
 where ${ \mathbb D}$, $Y$ and $K^A=(K^a,S^\a_i,\bar S_\ad^i)$ 
 are respectively the dilatation, $\sU(1)_R$, 
 special conformal and $S$-supersymmetry generators of the  
 $\cN=2$ superconformal algebra.
The nilpotency constraints \eqref{3.4} turn into 
\begin{subequations} \label{2.14-conf}
\bea
X^2&=&0~, \\
X \de_A \de_B X &=&0~,\\
X \de_A \de_B \de_C X &=&0~.
\eea
\end{subequations}
As in $\sSU(2)$ superspace,
$\D X$ is required to be nowhere vanishing, 
$\D X\neq 0$, where the covariantly antichiral projection operator is
\bea
{\D}:=\frac{1}{48}\de^{ij}\de_{ij}=-\frac{1}{48}\de^{\a\b}\de_{\a\b}~,~~~
\de_{ij}:=\de^\g_{(i}\de_{\g j)}~,~~~
\de_{\a\b}:=\de^k_{(\a}\de_{\b)k}
~,
\eea
and this expression for $\D$ is much simpler than the same operator in $\sSU(2)$ 
superspace, eq. \eqref{chiral-pr}.
For every primary superfield $U$ with the properties 
${ \mathbb D} U=0$,
$Y U=0$ and  $K^A U=0$,  
$\D U$ proves to be an antichiral primary superfield of dimension 2, 
that is: ${\mathbb D} \D U=2\D U$, $Y \D U=4\D U$, $K^A \D U=0$ and $\de_\a^i\D U=0$. An example of a superfield $U$ is provided by $X$. 
The Goldstino action \eqref{3.8} is uniquely lifted to conformal superspace to describe the dynamics of $X$.

As in the flat case, the nilpotent chiral superfield $X$ contains two independent component fields which can be 
identified with the $\q$-independent  components of the descendants 
$\c^\a_i := -\frac{1}{12} \de^{\a j}\de_{ij} X$ and $\D X$.
Unlike $X$ and $\D X$, the spinor superfield $\c^\a_i$  is not primary. 
It has two obvious properties:
\begin{subequations}
\bea
X \c^\a_i &=&0
~,
\\
\de_\b^j \c^\a_i &=&\d^\a_\b \d_i^j \D X
~.
\eea
It is much more difficult to derive the following relation 
\bea
\deb^\bd_{ j}\c^\a_{ i}
&=&
-\ri(\D X)^{-1}\Big(
\ve_{ij}\c^{\a k}\de^{\g\bd}\c_{\g k}
+\ve_{ij}\c^k_{\g}\de^{\g\bd}\c^\a_{ k}
+2\c^\g_{(i}\de^{\a\bd}\c_{\g j)}
\Big)
\non\\
&&
+(\D X)^{-2}\Big(\,
\ri\c_{ij}\de^{\a\bd}\D X
-\ri\ve_{ij}\,\c^{\a\g}\de_\g{}^{\bd}\D X
+\frac{2}{3}\ve_{ij}(\bar{\nabla}_\gd^{k} \bar{W}^{\bd\gd})\c_{kl}\c^{\a l}
\Big)
\non\\
&&
-\frac{4\ri}{3}\ve_{ij}(\D X)^{-3}\,\bar{W}^{\bd\gd} \c_{kl}\c^{\g k}\de^\a{}_{\gd}\c_{\g}^l
+3\ri\ve_{ij}(\D X)^{-4}\,\bar{W}^{\bd\gd} \c^4\de^\a{}_{\gd}\D X
~,
~~~~~~
\label{deb-chi-sugra}
\eea
\end{subequations}
which involves the super-Weyl tensor.
To completely specify the properties of $\c^\a_i$,  
we also need its $S$-supersymmetry transformations
\bea
\bar{S}_\bd^j\c^\a_i=0~,~~~
S^\b_j\c^\a_i
=
2(\D X)^{-1}\Big(
\ve^{\a\b}\c_{ij}
+\ve_{ij}\c^{\a\b}
\Big)~.
\eea
As in the flat-superspace case,  we make use of the 
definitions: $\c^{ij} :=\c^{\a i} \c^j_\a$, 
 $\c_{\a\b}:=\c^k_\a\c_{k \b}$ 
and $\c^4:=\frac{1}{3}\c^{ij}\c_{ij}=-\frac{1}{3}\c^{\a\b}\c_{\a\b}$.

Making use of the constraints \eqref{2.14-conf}, it is possible to prove that
 $X$ is  a composite superfield constructed from   $\c^\a_{ i}$,
\bea
X = \frac{\c^4 } {(\D X)^3}~, \qquad  \D X = \frac14 \de_\a^i \c^\a_i ~.
\label{repres-conf}
\eea
Using this representation, 
the nilpotency conditions \eqref{2.14-conf} are satisfied identically.

In the super-Poincar\'e case, the spinor Goldstino superfield $\X^\a_i$
was constructed from $X$ according to 
\eqref{Xi}. 
In the supergravity framework, we make use of a similar definition, 
\bea
\Xi^\a_{ i}:=f\frac{\c^\a_{i}}{\D X}
~,
\label{4.7}
\eea
with $\c^\a_i$ and $\D X$ given above.
 The superfield \eqref{4.7}  proves to satisfy the following constraints:
\bsubeq\label{deXi-sugra}
\bea
\de_\b^j\Xi^\a_{ i}
&=&
f\d_\b^\a\d^j_i
~,
\\
\deb^\bd_{ j}\Xi^\a_{i}
&=&
2\ri f^{-1}\Xi_{\g j}\de^{\g\bd}\Xi^\a_{ i}
-\ri\ve_{ij}f^{-3}(\de^\a{}_{\gd} \bar{W}^{\bd\gd})\Xi^4
-\ri\ve_{ij}f^{-3}\bar{W}^{\bd\gd}\de^\a{}_{\gd}\Xi^4
\non\\
&&
-2\ri\ve_{ij}f^{-3}\bar{W}^{\bd\gd} \Xi_{kl}\Xi^{\a k}\de_{\g\gd}\Xi^{\g l}
-\frac{4\ri}{3}f^{-3}\bar{W}^{\bd\gd} \Xi_{k(i} \Xi^{\a k}\de_{\g\gd}\Xi^\g_{j)}
\non\\
&&
-\frac{1}{3}\ve_{ij}f^{-2}(\deb_{\gd}^{k} \bar{W}^{\bd\gd} )\Xi_{kl}\Xi^{\a l}
-\frac{2}{3}f^{-2}(\deb_{\gd(i} \bar{W}^{\bd\gd} )\Xi_{j)k}\Xi^{\a k}
~,
\eea
\esubeq
which are the curved-superspace generalisation of 
the constraints \eqref{2.3a} and \eqref{2.3b}.

As with $\chi^\a_i$, $\Xi^\a_i$ is not primary. Its superconformal properties 
are determined by the relations 
\bea
S^\b_j\Xi^\a_i
=
2\ve^{\a\b}f^{-1}\Xi_{ij}
+2\ve_{ij}f^{-1}\Xi^{\a\b}
~,~~~~~~
\bar{S}_\bd^j\Xi^\a_i=0
~.
\eea
On the other hand, the composites $\c^4$ and $\Xi^4$
turn out to be primary superfields, 
\bea
K^A\c^4=0~,~~~~~~ 
K^A\Xi^4=0
~.
\eea
An important property of $\Xi^4$ is
\bea
\deb^\ad_{ i}\Xi^4
&=&
-2\ri\Xi^4\de^{\g\ad}\Xi_{\g i}
~.
\label{debXi^4}
\eea
This relation can be  used to check  that $X=\Xi^4\D X$ is chiral.
The same relation is useful to show that the dimensionless primary chiral scalar
\bea
\F:=\frac{1}{f^7}W^{-2}\bar\D (W^2\bar W^2\Xi^4\bar\Xi^4)
=\frac{1}{f^7}\bar\D (\bar W^2\Xi^4\bar\Xi^4)
\label{FfromXi-sugra}
\eea
has the following properties: 
\begin{subequations} \label{2.8_2.9-conf}
\bea
&\F^2=0~, ~~~
\F \de_A \de_B \F =0~,~~~
\F \de_A \de_B \de_C \F =0~,
\\
&f \F =W^{-2}\,  \F \bar \D \bar \F
~.
\eea
\esubeq
This primary chiral scalar is the unique extension of the irreducible Goldstino superfield
\eqref{3.12} to conformal superspace.
It is worth pointing out that \eqref{debXi^4} implies that $\F$ defined by \eqref{FfromXi-sugra} is proportional to $\Xi^4$.

The action for the Goldstino superfield $\X^\a_i$ coupled to supergravity 
is given by 
\bea
S_{\rm Goldstino} =- \frac{1}{f^6} \int \rd^4 x  \rd^4 {\q} \rd^4 \bar \q \,E\, W^2\bar W^2\,\X^4 \bar \X^4~.
\label{XiGoldstinoAction-sugra}
\eea
It can be recast in the form \eqref{3.13} if we make use of \eqref{FfromXi-sugra}.

It is important to observe that the constraints
\eqref{deXi-sugra} allow for the following unitary gauge condition
\bea
\Xi^\a_{ i}|_{\q=0} =0
~,
\eea
which completely fixes the local $Q$-supersymmetry invariance.
We now evaluate the Goldstino action \eqref{XiGoldstinoAction-sugra}
in this gauge  and show that it generates a positive contribution 
to the cosmological constant upon imposing standard superconformal gauge conditions.

First of all, we recall that any action given by an integral over the full superspace 
can equivalently be represented as an integral over the chiral subspace, 
\be 
\int \rd^4 x \rd^4 \q  \rd^4 \bar{\q} \,E \,\cL = \int \rd^4 x \rd^4 \q   \,\cE \, 
\bar\D \cL \ .
\ee
Next, reducing the chiral action to components gives
\begin{align}
S=& \int \rd^4x \,e \,\Big(\D +\cdots\Big)
\bar\D \cL \Big|_{\q=0}
~. 
\label{CA}
\end{align}
Here the ellipsis denotes terms involving supergravity fields and at most 
three spinor derivatives (see  \cite{Butter4DN=2,BN} for the 
complete expression). In the
\emph{unitary gauge},
it is easy to see that the component reduction of 
the action \eqref{XiGoldstinoAction-sugra} 
is
\begin{align}
S_{\rm Goldstino}=& 
-\frac{1}{f^6}\int \rd^4x \,e \,W^2\bar W^2(\D\Xi^4)(\bar\D\bar\Xi^4)\Big|_{\q=0}
=-f^2\int \rd^4x \,e \,W^2\bar W^2 \Big|_{\q=0}
~. 
\label{CA-2}
\end{align}
Here we have used the fact that in the unitary gauge we have
\bea
\de_\b^j\Xi^\a_{ i}|_{\q=0}=f\d_\b^\a\d^j_i
~,~~~~~~
\deb^\bd_{j}\Xi^\a_{i}|_{\q=0}
&=&
0~,
\eea
together with $\D\Xi^4|_{\q=0}=f^4$.
We also have to fix the local dilatation, $\sU(1)_R$ and superconformal ($K^A$) symmetries in a standard way \cite{deWPV} in order to end up with the canonically 
normalised Einstein-Hilbert action. In superspace this requires choosing 
the gauge $W =1$. The final expression for the cosmological constant proves to be
\bea
\L =f^2 -3\frac{m^2}{\k^2}~.
\eea
Here the second term on the right comes from the supersymmetric cosmological term
in the supergravity action \eqref{3.15}. 

Let us conclude this section with a few comments.
As mentioned above,  $\Xi^\a_i$ is not a primary superfield.
However, with the aid, for instance, of the  chiral  compensator $W$,
a primary extension of $\Xi^\a_i$ can be constructed. It turns out that the superfield 
\bea
{\bm \Xi}^\a_{i} =-\frac{1}{12W\D X}\Big(\de^{\a j}-3W^{-1}(\de^{\a j}W)\Big)
\de_{ij}(W X)
\eea
is primary.
Its evaluation gives
\bea
{\bm \Xi}^\a_{i}
&=&
\Xi^\a_i
-\frac{1}{2W}(\de^{\a j}W)\Xi_{ij}
-\frac{1}{2W}(\de_{\b i}W)\Xi^{\a\b}
\non\\
&&
-\frac{1}{6W}\Big((\de_{ij}W)-W^{-1}(\de^{\g}_{(i}W)(\de_{\g j)}W)\Big)\Xi^{jk}\Xi^\a_{k}
\non\\
&&
+\frac{1}{6W}\Big((\de^{\a\b}W)-3W^{-1}(\de^{(\a k}W)(\de^{\b)}_kW)\Big)\Xi_{ij}\Xi_{\b}^{ j}
\non\\
&&
-\frac{1}{12 W}\Big(
(\de^{\a j}\de_{ij}W)
-3W^{-2}(\de^{\a j}W)(\de_{ij}W)
\Big)
\Xi^4
~.
\label{primary-Xi}
\eea
In the derivation of \eqref{primary-Xi}, we have only used the fact that 
$W$ is a chiral primary superfield.
The Bianchi identity 
$\de^{ij}W=\deb^{ij}\bar{W}$ has not been used at all, and therefore the above construction does not require $W$ to be the field strength of a vector multiplet.

Multiplying \eqref{primary-Xi}
by $\D X$ gives  a primary extension of $\c^\a_i$, 
 \bea
 {\bm \c}^{\a}_i=f^{-1}{\bm \Xi}^{\a}_i\D X~.
 \eea
It holds that 
\bea
\Xi^4
={\bm \Xi}^4~, \qquad
\c^4={\bm \c}^4~.
\eea
For this reason the field redefinition $ { \X}^{\a}_i \to  {\bm \X}^{\a}_i$
does not affect any models constructed in terms of   $X$ or $\Xi^4$.


\section{Generalisations}

In conclusion we consider two generalisations inspired by the discussion in this paper.

\subsection{$\cN$-extended case}

Whilst the results in equations (\ref{2.3}) were derived in \cite{KM} with the case of ${\cal N} = 2$ supersymmetry in mind, they apply for arbitrary $\cN$-extended supersymmetry in four spacetime dimensions.\footnote{The $\cN=1$ case was also considered in \cite{KM}.} 
This is because they are derived from the coset parametrisation (\ref{2.1}) using only the anti-commutator 
\be
\{ Q_{\a}^i, \bar{Q}_{\ad \, j} \} = 2 P_{\a \ad} \delta^i_j~, \qquad i,j = 1, \dots , \cN
\ee and the conjugation rule $Q_{\a}^i{}^\dagger = \bar{Q}_{\ad i},$ which  are still applicable regardless of the range of the 
index $i$. However, the chiral action (\ref{2.4}) is specific to the ${\cal N} = 2$ case, and in the general $\cN$-extended 
case it must be replaced by 
\bea 
S =- \frac{1}{2f^{2(\cN-1)}} \int \rd^4 x  \rd^{2\cN} {\q} \,\J^{2\cN}
-\frac{1}{2f^{2(\cN-1)}}\int \rd^4 x  \rd^{2\cN} \bar{\q} \,\bar \J^{2\cN} ~,
\eea
where we have introduced the chiral scalar 
\bea
 \J^{2\cN} = \frac{2^\cN}{\cN! (\cN+1)!}\J_{i_1 j_1} \cdots \J_{i_\cN j_\cN}  
\ve^{i_1 \cdots i_\cN} \, \ve^{j_1 \cdots j_\cN} = \frac{2^\cN}{ (\cN+1)!}\, \det (\J_{ij})~,
\label{5.3}
\eea 
and as earlier, $\J_{ij}:= \J^{\a}_i \J_{\a j}$. The normalisation of the composite superfield  $ \J^{2\cN} $  is chosen so that 
$ \J^{2\cN} = \J_{11} \J_{22} \dots \J_{\cN \cN}$.
The determinant form of the $\cN$-extended chiral Lagrangian, eq. \eqref{5.3}, 
makes it analogous to the Volkov-Akulov theory \cite{VA,AV}.

\subsection{Generalisation of the Lindstr\"om-Ro\v{c}ek construction}

Lindstr\"om and Ro\v{c}ek \cite{LR} proposed to describe the Goldstino using 
a real scalar superfield $V$, which is nilpotent, 
$V^2=0$, 
and obeys the nonlinear constraint
\bea
f \bar S_0 S_0 V =\frac{1}{16} V \cD^\a (\bar \cD^2 -4R ) \cD_\a V~,
\label{6.1}
\eea
with $S_0$ the chiral compensator, $\bar \cD_\ad S_0 =0$, 
for the old minimal formulation for $\cN=1$ supergravity
\cite{WZ,old}.\footnote{Here we use the notation $S_0$ for the chiral compensator
following \cite{KU,Ferrara:1983dh}. In the superspace literature reviewed in \cite{GGRS},  it is usually denoted $\F$. The super-Weyl gauge
$S_0=1$ was used in \cite{LR}.} 
Actually,   $V$ was realised in  \cite{LR} only as a composite superfield, 
\bea
f \bar S_0 S_0 V = \bar \f \f~,
\label{6.2}
\eea 
constructed from the covariantly chiral scalar Goldstino superfield $\f$ 
(which is the curved-superspace extension of Ro\v{c}ek's nilpotent superfield
\cite{Rocek}) constrained by
\bea \label{X2g}
\bar \cD_{\dot\alpha} \f =0\,,\qquad  \f^2=0\,,\qquad 
{f} S_0^2 \f = -\frac 14  \f ({\bar \cD}^2 -4R)\bar  \f  \, ,
\eea
compare with \eqref{1.1}.
If instead $V$ is viewed as a fundamental Goldstino superfield, then 
it has been shown  \cite{BHKMS} that one has to impose 
the three nilpotency constraints 
\begin{subequations} \label{6.4}
\bea
V^2&=&0~, \\
V \cD_A \cD_B V &=&0~,\\
V \cD_A \cD_B \cD_C V &=&0~, 
\eea
\end{subequations}
in addition to \eqref{6.1}. It is also necessary to require that the descendant 
$\cD^\a W_\a$ is nowhere vanishing, where
\bea
W_\a = -\frac{1}{4} (\bar \cD^2 - 4R) \cD_\a V~.
\eea
The constraints \eqref{6.1} and \eqref{6.4} guarantee that $V$ contains 
a single independent component field -- the Goldstino, which is the lowest 
($\q$-independent) component of $W_\a$.
The Goldstino action is
\bea
S_{\rm Goldstino} = -f  \int \rd^4 x \rd^2 \q  \rd^2 \bar{\q} \, E\,\bar S_0 S_0 V~.
\label{6.6}
\eea

The constraints \eqref{6.1} and \eqref{6.4}, as well as the action \eqref{6.6}, are 
invariant under super-Weyl transformations \cite{HT} of the form
\begin{subequations} 
\label{superweyl}
\bea
\d_\s \cD_\a &=& ( {\bar \s} - \hf \s)  \cD_\a + (\cD^\b \s) \, M_{\a \b}  ~, \\
\d_\s \bar \cD_\ad & = & (  \s -  \hf {\bar \s})
\bar \cD_\ad +  ( \bar \cD^\bd  {\bar \s} )  {\bar M}_{\ad \bd} ~,\\
\d_\s \cD_{\a\ad} &=& \hf( \s +\bar \s) \cD_{\a\ad} 
+\frac{\ri}{2} (\bar \cD_\ad \bar \s) \cD_\a + \frac{\ri}{2} ( \cD_\a  \s) \bar \cD_\ad \non \\
&& + (\cD^\b{}_\ad \s) M_{\a\b} + (\cD_\a{}^\bd \bar \s) \bar M_{\ad \bd}~,
\eea
\end{subequations}
where $\s$ is an arbitrary covariantly chiral scalar superfield,  $\bar \cD_\ad \s =0$. 
It is assumed that $V$ is super-Weyl inert, while $S_0 $ transforms 
as $\d_\s S_0 = \s S_0$.

It is possible to follow a different path than the one just discussed, 
in the spirit of the nilpotent  $\cN=1$ chiral construction of \cite{Casalbuoni,KS}.
Specifically,  we consider a Goldstino superfield $V$ which only obeys the nilpotency constraints \eqref{6.4}, in conjunction with the requirement 
that $\cD^\a W_\a$ is nowhere vanishing. It has two independent component fields, 
the Goldstino $W_\a|_{\q=0}$ and the auxiliary scalar $\cD^\a W_\a|_{\q=0}$. 
One may show that the constraints \eqref{6.4} 
imply the representation\footnote{In the case that $V$ obeys the constraint \eqref{6.1}, 
the relation \eqref{6.8} reduces to $(f\bar S_0 S_0)^3 V = 16W^2 \bar W^2 $, which was derived in \cite{BHKMS} in the flat-superspace limit.}
\bea
V = - 4 \frac{W^2 \bar W^2}{(\cD^\a W_\a)^3}~, \qquad 
W^2 = W^\a W_\a~,
\label{6.8}
\eea
which ensures \eqref{6.4} is  identically satisfied.
The relation \eqref{6.8} is super-Weyl invariant, 
since $W_\a$ and $\cD^\a W_\a$ transform as super-Weyl primary 
superfields, $\d_\s W_\a = \frac{3}{2} \s W_\a$ and $\d_\s (\cD^\a W_\a)
= (\s+\bar \s) \cD^\a W_\a$.

The constraints \eqref{6.4} are invariant under local re-scalings of $V$
\bea
V ~ \to ~ \re^\t V~,
\eea
with $\t$ an arbitrary real scalar superfield.
The dynamics of this supermultiplet is governed by the action
\bea
\widetilde{S}_{\rm Goldstino} =   \int \rd^4 x \rd^2 \q  \rd^2 \bar{\q} \, E\,\Big\{
\frac{1}{16} V \cD^\a (\bar \cD^2 -4R ) \cD_\a V- 2f \bar S_0 S_0 V\Big\}~.
\label{6.10}
\eea
Varying the Goldstino superfield according to $\d V = \t V$, with $\t$ being arbitrary, gives the constraint \eqref{6.1} as the corresponding equation of motion. 
Then the action \eqref{6.10} reduces to \eqref{6.6}.

Within the new minimal formulation for $\cN=1$ supergravity \cite{new}, the compensator is a  real covariantly linear scalar superfield,
\bea  
(\bar \cD^2 -4R) {\mathbb L}  =0~, \qquad \bar {\mathbb L}= {\mathbb L}~,
\eea
with the super-Weyl transformation $\d_\s {\mathbb L} = (\s+\bar \s) {\mathbb L}$,
see \cite{BK,Ferrara:1983dh,GGRS} for reviews.
The action for supergravity coupled to the Goldstino superfield $V$ is 
\bea
S=   \int \rd^4 x \rd^2 \q  \rd^2 \bar{\q} \, E\, \left\{\frac{3}{\k^2}
{\Bbb L}\, {\rm ln} \frac{\Bbb L}{|S_0|^2}
+\frac{1}{16} V \cD^\a (\bar \cD^2 -4R ) \cD_\a V- 2f {\mathbb L} V
 \right\}~,
 \label{5.15}
\eea
where now $S_0$ is a purely gauge degree of freedom. 

If $V$ is  a real unconstrained superfield, the action
\eqref{5.15} describes new minimal supergravity coupled to 
 a massless  vector supermultiplet 
with a Fayet-Iliopoulos term (see, e.g., \cite{Ferrara:1983dh}).
The action is invariant under $\sU(1) $ gauge 
transformations $\d V = \l + \bar \l$, with the gauge parameter $\l$  being chiral,
$\bar \cD_\ad \l =0$.  
However, in our case $V$ is subject to the nilpotency conditions 
\eqref{6.4}, which are incompatible with the gauge invariance.
These nilpotency conditions guarantee that the Goldstino and the auxiliary field 
are the only independent component fields of $V$. 

An important feature of unbroken new minimal supergravity is that it does not allow 
any supersymmetric cosmological term \cite{SohniusW2,SohniusW3}.\footnote{Among the known off-shell 
formulations for $\cN=1$ (see \cite{BK,GGRS} for reviews), supersymmetric cosmological terms 
exist only for the old minimal supergravity \cite{FvN2,Siegel78} 
and the $n=-1$ non-minimal supergravity as formulated in \cite{ButterK12}.}
Our  action for spontaneously broken supergravity \eqref{5.15} leads, at the component level, to
a positive cosmological constant which is generated by the Goldstino superfield. 
The cosmological constant in  \eqref{5.15} is strictly positive since there 
is no supersymmetric cosmological term producing a negative contribution.

$~$\\
\noindent
{\bf Acknowledgements:}\\
SMK is grateful to Gianguido Dall'Agata, Renata Kallosh and Paul Townsend for discussions, and to Dima Sorokin for comments on the manuscript.
SMK and GT-M thank the Galileo Galilei Institute for Theoretical Physics for hospitality 
and the INFN for partial support at the initial stage of this project.
GT-M is grateful to the School of Physics at the University of Western Australia for kind hospitality
and support during part of this project.
The work of SMK and GT-M was supported in part by the Australian 
Research Council,  project No. DP140103925.
The work of GT-M is supported by the Interuniversity Attraction Poles Programme
initiated by the Belgian Science Policy (P7/37), and in part by COST Action MP1210 
``The String Theory Universe.''


\appendix

\section{The component content of \mbox{$\bm{\F}$}}
\label{Appendix0} 

In this appendix we elaborate on the component content of the 
chiral scalar Goldstino superfield
defined by the constraints \eqref{2.8} and \eqref{2.9}.
For this it is relevant to introduce the two descendants of $\F$:
\bea\label{def-chi-F}
\c^\a_i
:= -\frac{1}{12} D^{\a j}D_{ij}\F
= -\frac{1}{12} D_{\b i}D^{\a\b} \F
~,~~~~~~
F:=D^4\F
~.
\eea
The Goldstini may be identified with $\c^\a_i|_{\q=0}$.
By assumption, the field $F|_{\q=0}$ is nowhere vanishing.
The constraints \eqref{2.8} prove to imply the  relations:
\bsubeq\label{Phi-from-chi-and-F}
\bea
D_\a^i\F
&=&
\frac{4}{3}\c_{\a j}\c^{ij}F^{-2}
= \frac{4}{3} \c^{\b i} \c_{\a \b }  F^{-2}
~,
\\
D^{ij}\F &=&
-4\c^{ij}F^{-1}~,   \label{A.2b}\\
D_{\a\b}\F &=&
4\c_{\a\b}F^{-1}
~,\\
\F&=&\c^4F^{-3} \label{A.2d}
~,
\eea
\esubeq
where we have introduced the composites
\bea
\c^{ij} =\c^{\a i} \c_\a^j=\c^{ji}
~,~~~
\c_{\a\b} =\c_\a^{i} \c_{\b i}=\c_{\b\a} 
~,~~~
\c^4:=\frac{1}{3}\c^{ij}\c_{ij}=-\frac{1}{3}\c^{\a\b}\c_{\a\b}
~.~~~~~~
\eea
The relations 
\eqref{Phi-from-chi-and-F} 
imply that all the components of $\F$ are expressed in terms of 
$\c^\a_i|_{\q=0}$ and $F|_{\q=0}$.
Furthermore, by applying the operator $D^4$ to both sides of \eqref{2.9} one can derive the following nonlinear equation on $F$ and its conjugate
\bea
f F
&=&
-2\ri\c^\a_i\pa_{\a\ad}\bar\c^{\ad i}
+F \bar F
+\frac{\c^{ij}}{F}\Box\frac{\bar\c_{ij}}{\bar F}
-\frac{\c^{\a\b}}{F}\pa_{\a\ad}\pa_{\b\bd}\frac{\bar\c^{\ad\bd}}{\bar F}
\non\\
&&
-\frac{8\ri}{9}\frac{\c^{ij}\c^\a_{j}}{F^{2}}\Box\pa_{\a\ad}
\frac{\bar\c_{ik}\bar\c^{\ad k}}{\bar F^{2}}
+\frac{\c^4}{F^3} \Box^2\frac{\bar\c^4}{\bar{F}^3}
~.
\eea
This equation can be uniquely solved by iteration in order to express $F$ in terms of 
$\c^\a_i$ and its complex conjugate 
$\bar\c^{\ad i}:=\overline{(\c^\a_i)}$, 
\bea
F = f - \frac{2\ri }{f} (\pa_{\a\ad} \c^\a_i) \bar \c^{\ad i} + O(\c^4)~.
\eea
 The series terminates since $\c^\a_i$ and $\bar\c^{\ad i}$ are anti-commuting.


\section{\mbox{$\bm{\sSU(2)}$} superspace} \label{AppendixA}

This appendix contains a summary of the formulation for $\cN=2$ conformal supergravity  \cite{KLRT-M1} in $\sSU(2)$ 
superspace \cite{Grimm}.
A curved $\cN = 2$ superspace
is parametrised by local coordinates 
$z^M = (x^m, \theta^\mu_\imath, \bar{\theta}_{\dot{\mu}}^\imath 
= \overline{\theta_{\mu \imath} } \,)$, where $m = 0, 1, 2, 3$ and
$\mu, \dot \m, \imath = 1, 2$.
The superspace structure group is chosen to be $\sSL(2, \dsC) \times \sSU(2)$, 
and the covariant derivatives $\cD_A = (\cD_a, \cD_\a^i, \bar \cD^\ad_i)$ read
\bea
\cD_A &=& E_A + \Phi_A{}^{kl} J_{kl}+ \hf \Omega_A{}^{bc} M_{bc} \non \\
		  &=& E_A + \Phi_A{}^{kl} J_{kl}+ \Omega_A{}^{\b\g} M_{\b\g} 
		  + \bar{\Omega}_A{}^{ \dot{\b} \dot{\g} } \bar M_{\dot{\b}\dot{\g}}~.
\eea
Here $E_A =E_A{}^M \pa_M$, 
$M_{cd}$ and $J_{kl}$ are the generators of the Lorentz and $\sSU(2)$ groups respectively, 
and $\O_A{}^{bc}$ and $\Phi_A{}^{kl}$ the corresponding connections. 
The action of the generators on the covariant derivatives are defined as:
\bea
[M_{\a\b},  \cD_\g^i] &=& \ve_{\g (\a } \cD_{\b)}^i 
\ , \qquad 
\big[ J_{kl}, \cD_\a^i \big] = - \d^i_{(k} \cD_{\a l)} 
\ ,
\label{Lorentz-SU(2)-Gen}
\eea
together with their complex conjugates.

The algebra of covariant derivatives is \cite{KLRT-M1}
\begin{subequations} \label{A.3}
\bea
\{\cD_\a^i,\cD_\b^j\}&=&\phantom{+}
4S^{ij}M_{\a\b}
+2\ve^{ij}\ve_{\a\b}Y^{\g\d}M_{\g\d}
+2\ve^{ij}\ve_{\a\b}\bar{W}^{\gd\dd}\bar{M}_{\gd\dd}
\non\\
&&
+2 \ve_{\a\b}\ve^{ij}S^{kl}J_{kl}
+4 Y_{\a\b}J^{ij}~,
\label{acr1} \\
\{\cD_\a^i,\cDB^\bd_j\}&=&
-2\ri\d^i_j(\s^c)_\a{}^\bd\cD_c
+4\d^{i}_{j}G^{\d\bd}M_{\a\d}
+4\d^{i}_{j}G_{\a\gd}\bar{M}^{\gd\bd}
+8 G_\a{}^\bd J^{i}{}_{j}~,
\label{acr3} 
\eea
\end{subequations}
together with the complex conjugate of \eqref{acr1},
see \cite{KLRT-M1} for the explicit expressions for the commutators
${[}\cD_a,\cD_\b^j{]}$ and ${[}\cD_a,\cDB^\bd_j{]}$.
Here the real four-vector $G_{\a \ad} $,
the complex symmetric  tensors $S^{ij}=S^{ji}$, $W_{\a\b}=W_{\b\a}$, 
$Y_{\a\b}=Y_{\b\a}$ and their complex conjugates 
$\bar{S}_{ij}:=\overline{S^{ij}}$, $\bar{W}_{\ad\bd}:=\overline{W_{\a\b}}$,
$\bar{Y}_{\ad\bd}:=\overline{Y_{\a\b}}$ 
are constrained by 
the Bianchi identities \cite{Grimm,KLRT-M1}.
   The latter  comprise 
 the  dimension-3/2 identities 
\begin{subequations}\label{A.5}
\bea
\cD_{\a}^{(i}S^{jk)}= {\bar \cD}_{\ad}^{(i}S^{jk)}&=&0
~,~~
\cDB^\ad_iW_{\b\g}=0~,~~
\cD_{(\a}^{i}Y_{\b\g)}=0~,~~
\cD_{\a}^{i}S_{ij}+\cD^{\b}_{j}Y_{\b\a}=0~,
~~~~~~~~~
\\
\cD_\a^iG_{\b\bd}&=&
-{\frac14}\cDB_{\bd}^iY_{\a\b}
+\frac{1}{12}\ve_{\a\b}\cDB_{\bd j}S^{ij}
-{\frac14}\ve_{\a\b}\cDB^{\gd i}\bar{W}_{\bd\gd}~,
\eea
\end{subequations}
together with their complex conjugates
as well as the dimension-2 relation
\bea
\big( \cD^i_{(\a} \cD_{\b) i}
-4Y_{\a\b} \big) W^{\a\b}
&=& \big( \cDB_i^{( \ad }\cDB^{\bd ) i}
-4\bar{Y}^{\ad\bd} \big) \bar{W}_{\ad\bd}
~.
\label{dim-2-constr}
\eea


\section{Conformal superspace} \label{AppendixConfSuperspace}

This appendix contains a summary of the formulation for $\cN=2$ conformal supergravity  
in conformal  
superspace \cite{Butter4DN=2} employed in section \ref{section4}.
We use the notations of \cite{BN} which are consistent with those of \cite{KLRT-M1} and 
Appendix \ref{AppendixA} and review the results necessary for deriving results in section \ref{section4}.
The structure group of $\cN=2$ conformal superspace 
is chosen to be $\sSU(2, 2|2)$
and the covariant derivatives  
$\nabla_A = (\nabla_a, \nabla_\a^i , \bar{\nabla}^\ad_i)$ have the form
\begin{align} \nabla_A &= E_A + \hf \Omega_A{}^{ab} M_{ab} + \Phi_A{}^{ij} J_{ij} + \ri \Phi_A Y 
+ B_A \mathbb{D} + \frak{F}_{A}{}^B K_B \non\\
&= E_A + \Omega_A{}^{\b\g} M_{\b\g} + \bar{\Omega}_A{}^{\bd\gd} \bar{M}_{\bd\gd} + \Phi_A{}^{ij} J_{ij} + \ri \Phi_A Y 
+ B_A \mathbb{D} + \frak{F}_{A}{}^B K_B \ .
\end{align}
Here, as in $\sSU(2)$ superspace,
$E_A =E_A{}^M \pa_M$, $M_{cd}$ and $J_{kl}$ are the generators of the Lorentz and $\sSU(2)$ $R$-symmetry groups 
respectively, and $\O_A{}^{bc}$ and $\Phi_A{}^{kl}$ the corresponding connections. 
The remaining generators and corresponding connections are: 
$Y$  and $\Phi_A$ for the $\sU(1)$ $R$-symmetry group; 
$\mathbb D$ and $B_A$ for the dilatations;
 $K^A = (K^a, S^\a_i, \bar{S}_\ad^i)$
and $\frak{F}_A{}^B$
for the special superconformal generators.

The Lorentz and $\sSU(2)$ generators act on $\de_A$ as in the $\sSU(2)$ superspace case, see eq. 
\eqref{Lorentz-SU(2)-Gen}.
The $\sU(1)_R$ and dilatation generators obey
\bsubeq
\bea
&{[}Y, \nabla_\a^i{]}= \nabla_\a^i ~,\quad {[}Y, \bar\nabla^\ad_i{]} = - \bar\nabla^\ad_i~,
\\
&{[}\mathbb{D}, \nabla_a{]} = \nabla_a ~,\quad
{[}\mathbb{D}, \nabla_\a^i{]} = \hf \nabla_\a^i ~,\quad
{[}\mathbb{D}, \bar\nabla^\ad_i{]} = \hf \bar\nabla^\ad_i ~.
\eea
\esubeq
The special superconformal generators $K^A$ transform 
under  Lorentz and $\sSU(2)$  as
\bea
&[M_{ab}, K_c] = 2 \eta_{c [a} K_{b]} ~, \quad
[M_{\a\b} , S^\g_i] =\d^\g_{(\a}S_{\b)i} ~, \quad
[J_{ij}, S^\g_k] = - \ve_{k (i} S^\g_{j)} ~,
\eea
together with their complex conjugates,
while their transformation under $\sU(1)$ and dilatations is:
\bsubeq
\bea
&[Y, S^\a_i] = - S^\a_i ~, \quad
[Y, \bar{S}^i_\ad] = \bar{S}^i_\ad~, \non \\
&[\mathbb{D}, K_a] = - K_a ~, \quad
[\mathbb{D}, S^\a_i] = - \hf S^\a_i ~, \quad
[\mathbb{D}, \bar{S}_\ad^i] = - \hf \bar{S}_\ad^i ~.
\eea
\esubeq
The generators $K^A$ obey
\begin{align}
\{ S^\a_i , \bar{S}^j_\ad \} &= 2 \ri \d^j_i (\s^a)^\a{}_\ad K_a
~,
\end{align}
while the nontrivial (anti-)commutators of the algebra of $K^A$ with $\nabla_B$ are given by
\bsubeq
\bea
&[K^a, \nabla_b] = 2 \delta^a_b \mathbb{D} + 2 M^{a}{}_b 
~,\non \\
&\{ S^\a_i , \nabla_\b^j \} = 
2 \d^j_i \d^\a_\b \mathbb{D} - 4 \d^j_i M^\a{}_\b 
- \d^j_i \d^\a_\b Y + 4 \d^\a_\b J_i{}^j ~,\non \\
&[K^a, \nabla_\b^j] = -\ri (\s^a)_\b{}^\bd \bar{S}_\bd^j \ , \quad
[S^\a_i , \nabla_b] = \ri (\s_b)^\a{}_\bd \bar{\nabla}^\bd_i 
\ ,
\eea
\esubeq
together with complex conjugates.

The  (anti-)commutation relations of the covariant derivatives $\de_A$ \cite{Butter4DN=2,BN}
relevant for calculations in this paper are
\begin{subequations}\label{CSGAlgebra}
\begin{align}
\{ \nabla_\a^i , \nabla_\b^j \} &= 2 \ve^{ij} \ve_{\a\b} \bar{W}_{\gd\dd} \bar{M}^{\gd\dd} + \hf \ve^{ij} \ve_{\a\b} \bar{\nabla}_{\gd k} \bar{W}^{\gd\dd} \bar{S}^k_\dd - \hf \ve^{ij} \ve_{\a\b} \nabla_{\g\dd} \bar{W}^\dd{}_\gd K^{\g \gd}~, \\
\{ \nabla_\a^i , \bar{\nabla}^\bd_j \} &= - 2 \ri \d_j^i \nabla_\a{}^\bd~, \\
[\nabla_{\a\ad} , \nabla_\b^i ] &= - \ri \ve_{\a\b} \bar{W}_{\ad\bd} \bar{\nabla}^{\bd i} - \frac{\ri}{2} \ve_{\a\b} \bar{\nabla}^{\bd i} \bar{W}_{\ad\bd} \mathbb{D} - \frac{\ri}{4} \ve_{\a\b} \bar{\nabla}^{\bd i} \bar{W}_{\ad\bd} Y 
+ \ri \ve_{\a\b} \bar{\nabla}^\bd_j \bar{W}_{\ad\bd} J^{ij}
	\eol & \quad
	- \ri \ve_{\a\b} \bar{\nabla}_\bd^i \bar{W}_{\gd\ad} \bar{M}^{\bd \gd} - \frac{\ri}{4} \ve_{\a\b} \bar{\nabla}_\ad^i \bar{\nabla}^\bd_k \bar{W}_{\bd\gd} \bar{S}^{\gd k} + \frac{1}{2} \ve_{\a\b} \nabla^{\g \bd} \bar{W}_{\ad\bd} S^i_\g
	\eol & \quad
	+ \frac{\ri}{4} \ve_{\a\b} \bar{\nabla}_\ad^i \nabla^\g{}_\gd \bar{W}^{\gd \bd} K_{\g \bd}~,
\end{align}
\end{subequations}
together with complex conjugates.
The superfield $W_{\a\b} = W_{\b\a}$ and its complex conjugate
${\bar{W}}_{\ad \bd} := \overline{W_{\a\b}}$ are dimension one conformal primaries,
$K_A W_{\a\b} = 0$, and obey the additional constraints
\begin{align}
\bar{\nabla}^\ad_i W_{\b\g} = 0~,\qquad
\nabla_{\a}^k \nabla_{\b k} W^{\a\b} &=\bar\nabla^{\ad}_k \bar\nabla^{\bd k}  \bar{W}_{\ad\bd} ~.
\end{align}


\section{Nilpotent \mbox{$\bm{\cN=1}$}  supergravity} \label{appendixD}

Consider $\cN=1$ supergravity coupled to a covariantly 
chiral scalar $\cX$, $\bar \cD_\ad \cX =0$,
subject to the nilpotency condition 
\bea
\cX^2 =0~.
\label{D.1}
\eea
The complete action, which is equivalent to the action used in \cite{BFKVP,HY}, is 
\bea
S &=& 
\int {\rm d}^{4} x \rd^2\q\rd^2\bar\q\,
E\, \Big(- \frac{3}{ \k^2} \bar S_0 S_0 + \bar \cX  \cX \Big) \non \\
&& \qquad \qquad \qquad 
+ \left\{  \int {\rm d}^{4} x \rd^2 \q\,
{\cal E} \,S_0^3 \Big(  \frac{ \m}{ \k^2} -f \frac{\cX}{S_0}  \Big)
+ {\rm c.c.} \right\}
~.
\label{D.2}
\eea
Under the super-Weyl transformation \eqref{superweyl}, $\cX$ 
transforms as a primary dimension-1 superfield, $\d_\s \cX = \s \cX$. 

Varying  \eqref{D.2} with respect to the chiral compensator $S_0$ gives 
the equation
\bea
{\mathbb R} -{\m} = -\frac{2}{3} f \k^2 \frac{\cX}{S_0} ~,
\label{D.3}
\eea
where we have introduced the super-Weyl invariant chiral scalar 
\bea
{\mathbb R} = -\frac{1}{4} S_0^{-2} (\bar \cD^2 - 4R) \bar S_0~.
\eea
Eq.\,\eqref{D.3} 
allows for two interpretations. 
Firstly, it is the equation of motion for $S_0$, with $\cX$ being a spectator superfield. 
Secondly, it allows us to express $\cX$ as a function of 
the supergravity fields. 
 The nilpotency constraint \eqref{D.1} and  the equation of motion 
\eqref{D.3} imply that the chiral curvature becomes  nilpotent \cite{K15,BMST}, 
\bea
({\mathbb R} -{\m})^2 =0~.
\label{D.5}
\eea
Making use of \eqref{D.3}  once more, 
the functional  \eqref{D.2} can be rewritten as a higher-derivative supergravity action \cite{K15}
\bea
S = \Big( \frac{3}{2f \k^2}\Big)^2  
\int {\rm d}^{4} x \rd^2\q\rd^2\bar\q\,
E \, \bar S_0 S_0|{\mathbb R} -{ \m} |^2
- \left\{  \hf  \frac{ \m}{ \k^2} 
\int {\rm d}^{4} x \rd^2 \q\,
{\cal E} \,S_0^3 
+ {\rm c.c.} \right\}
~,~~~
\label{D.6}
\eea
where $\mathbb R$ is subject to the constraint \eqref{D.5}.
This action does not involve  the Goldstino superfield explicitly.

The nilpotency condition \eqref{D.1} is preserved if $\cX$ is locally rescaled, 
\bea
\cX ~\to ~\re^{\t} \cX~, \qquad \bar \cD_\ad \t=0~.
\label{D.7}
\eea
Requiring the action \eqref{D.2} to be stationary under such re-scalings  of $\cX$ 
gives 
\bea
\cX = \f ~, 
\eea
where $\f$ is the Lindstr\"om-Ro\v{c}ek chiral scalar defined by \eqref{X2g}. If the compensator satisfies its equation 
of motion \eqref{D.3}, then the chiral curvature obeys the nonlinear constraint 
\bea
\frac{2}{3} (f \k)^2 S_0^2 ({\mathbb R} -{\m}) = \frac 14 ({\mathbb R} -{\m})
(\bar \cD^2 -4R ) \Big[ \bar S_0 ( \bar{\mathbb R} -{\m})\Big]~, 
\label{D.9}
\eea
in addition to the nilpotency condition \eqref{D.5}.
Making use of \eqref{D.9} turns the action \eqref{D.6} into 
\bea
S &=& - \frac{3}{ 2\k^2}
\int {\rm d}^{4} x \rd^2\q\rd^2\bar\q\,
E\,  \bar S_0 S_0 
+ \left\{ \frac{ \m}{4 \k^2} \int {\rm d}^{4} x \rd^2 \q\,
{\cal E} \,S_0^3
+ {\rm c.c.} \right\}
~.
\label{D.10}
\eea
This is a pure supergravity action with rescaled Newton's constant and cosmological 
parameter, $\k^2 \to 2\k^2$ and $\m \to \hf \m$. The chiral compensator $S_0$ in 
\eqref{D.10} obeys the nilpotency condition \eqref{D.5} and the nonlinear constraint 
\eqref{D.9}. In a super-Weyl gauge $S_0=1$, these conditions turn into 
\begin{subequations}
\bea
(R-\m)^2 &=&0~, \\
\frac{2}{3} (f \k)^2  ({R} -{\m}) &=& \frac 14 ({R} -{\m})
(\bar \cD^2 -4R )  ( \bar{ R} -{\m})~.
\eea
\end{subequations}

Other approaches to nilpotent $\cN=1$ supergravity 
were developed in \cite{ADFS,BMST,DFKS,AM,Cribiori:2016qif}.


\begin{footnotesize}

\end{footnotesize}

\end{document}